\documentclass[journal]{IEEEtran}
\usepackage{bm}
\usepackage{verbatim}
\usepackage{amsfonts}
\usepackage{amssymb}
\usepackage{stfloats}
\usepackage{cite}
\usepackage{graphicx}
\usepackage{psfrag}
\usepackage{subfigure}
\usepackage{amsmath}
\usepackage{array}
\usepackage{epstopdf}
\usepackage{authblk}
\usepackage{graphicx} 
\usepackage{amsthm}         
\usepackage{lipsum}
\usepackage{verbatim} 
\usepackage{authblk}
\usepackage{mathtools}
\usepackage{cuted}
\usepackage{authblk}
\usepackage{booktabs}
\usepackage{subfigure}
\usepackage{siunitx}
\usepackage{algorithm, algorithmic}
\usepackage{mathtools}
\usepackage{soul} 
\usepackage{xcolor}

\newtheorem{Lemma}{Lemma}

\newtheorem{Proposition}{Proposition}

\def\bmu{{\boldsymbol{\mu}}}
\def\b0{{\mathbf{0}}}
\def\bw{{\mathbf{w}}}
\def\bx{{\mathbf{x}}}
\def\bC{{\mathbf{C}}}
\def\bI{{\mathbf{I}}}
\def\cN{\mathcal{N}}
\def\cW{\mathcal{W}}
\def\l({\left(}
\def\r){\right)}

\begin{document}

\title{Knowledge-Based Ultra-Low-Latency Semantic Communications for Robotic Edge Intelligence}

\author{{Qunsong Zeng, Zhanwei Wang, You Zhou, Hai Wu, Lin Yang, and Kaibin Huang}
\thanks{Q. Zeng, Z. Wang, Y. Zhou, H. Wu, and K. Huang are with the Department of Electrical and Electronic Engineering, The University of Hong Kong, Hong Kong, China. L. Yang is with Noah's Ark Lab, Huawei, Shenzhen, China. Corresponding authors: K. Huang (Email: huangkb@eee.hku.hk) and L. Yang (yanglin33@huawei.com).}
}

\maketitle

\begin{abstract}
The \emph{sixth-generation} (6G) mobile networks will feature the widespread deployment of \emph{artificial intelligence} (AI) algorithms at the network edge, which provides a platform for supporting robotic edge intelligence systems.
In such a system, a large-scale \emph{knowledge graph} (KG) is operated at an edge server as a ``remote brain'' to guide remote robots on environmental exploration or task execution.
In this paper, we present a new air-interface framework targeting the said systems, called knowledge-based robotic \emph{semantic communications} (SemCom), which consists of a protocol and relevant transmission techniques. 
First, the proposed robotic SemCom protocol defines a sequence of system operations for executing a given robotic task. 
They include identification of all task-relevant \emph{knowledge paths} (KPs) on the KG, semantic matching between KG and object classifier, and uploading of robot's observations for objects recognition and feasible KP identification.
Next, to support \emph{ultra-low-latency (observation) feature transmission} (ULL-FT), we propose a novel transmission approach that exploits classifier's robustness, which is measured by \emph{classification margin}, to compensate for a high \emph{bit error probability} (BEP) resulting from ultra-low-latency transmission (e.g., short packet and/or no coding). 
By utilizing the tractable \emph{Gaussian mixture} (GM) model, we mathematically derive the relation between BEP and classification margin under constraints on classification accuracy and transmission latency.
The result sheds light on system requirements to support ULL-FT.
Furthermore, for the case where the classification margin is insufficient for coping with channel distortion, we enhance the ULL-FT approach by studying retransmission and multi-view classification for enlarging the margin and further quantifying corresponding requirements.
Finally, experiments using deep neural networks as classifier models and real datasets are conducted to demonstrate the effectiveness of ULL-FT in communication latency reduction while providing a guarantee on accurate feasible KP identification.
\end{abstract}

% \vspace{-2mm}
\section{Introduction}\label{Introduction}
The \emph{sixth-generation} (6G) mobile networks is set apart from its predecessors by the anticipated widespread implementation of machine learning and \emph{artificial intelligence} (AI) algorithms at the network edge~\cite{letaief2019roadmap}. 
This spread of computing from cloud to edge brings the benefits of streamlined processing of mobile data, protection of user privacy, relieved network traffic, and provisioning of ultra-low-latency access~\cite{park2019wireless}.  Among others, to connect machines to large-scale AI models at the edge is  one key mission of 6G to meet the ubiquity of AI demands for next generation \emph{Internet-of-Things} (IoT) applications~\cite{rong20216g}. This motivates us to propose in this work a framework of knowledge-based  robotic \emph{semantic communications} (SemCom) that comprises a knowledge based  communication protocol and enabling techniques.  The framework promises to provide ultra-low-latency, robotic-intelligence oriented  connectivity between a robot to an edge server that operates  a large-scale \emph{knowledge graph} (KG) to guide the robot on executing a given task.

Current research in edge AI focuses on efficient deployment of large-scale AI models, which are pre-trained on distributed mobile data, at the network edge to provide remote inference services to edge devices \cite{lim2020federated, wang2020convergence}. 
A popular system architecture, known as \emph{split inference}, partitions a pre-trained model into device and server sub-models, where the former is used for local feature extraction and the latter for on-server high-accuracy prediction on the uploaded features~\cite{shao2020communication}. 
The offloading of inference tasks to  servers allows devices to access large-scale AI models that are infeasible to execute on devices. To implement split inference, researchers have studied the optimization of the model-splitting point   according to different practical factors, including on-device computation resources, radio resources, and latency requirements,   to minimize communication overhead and accommodate the heterogeneity of devices' hardware~\cite{shi2019improving,li2019edge,shao2022task,huang2020dynamic}. 
In terms of  communication techniques, each  design for edge AI is  task-oriented and based on a chosen end-to-end metric (e.g., end-to-end latency, accuracy, or throughput) to achieve better  system performance and a higher system efficiency as opposed to the traditional approach that separates communication and computing~\cite{jankowski2020joint,wen2023task,lan2022progressive}. 
While the area of edge AI is relatively  new and targets generic IoT applications, in the area of robotics, the paradigm of using the cloud (or edge cloud) as a knowledge base to guide robots on  executing sophisticated manipulation tasks has been extensively investigated (see e.g.,~\cite{CL_KG, KG_robotic_ICRA, KG_robotic_IJRR}).  
In particular, the remote knowledge base helps robots to detect semantic knowledge from observations of the external  environments, facilitating  human-robot cooperation (see e.g.,~\cite{KG_robotic_IJRR}). 
On the other hand, the availability of real-time sensing data at robots can be exploited to continuously learn knowledge embedding via acquiring previously unknown concepts (see e.g.,~\cite{CL_KG}).  
However, the design of an efficient air interface for robotic edge AI systems remains as an open area. 
To maximize communication efficiency, the customized wireless techniques  aim at conveying meanings in human messages or  effective task execution. 

These principles were originally proposed by Warren Weaver in 1950s and now are at the core of an emerging area, called \emph{semantic communications}, which emphasizes transmitting semantically relevant information to achieve a higher communication efficiency~\cite{weaver1953recent,lan2021semantic}. From the perspective of networking, researchers have proposed the addition of a ``semantic-effectiveness plane'' to the existing layered network architecture to endow on radio-access techniques awareness of semantics and task effectiveness \cite{popovski2020semantic}. From the perspective of system designs, one focus is on the use of the so called \emph{joint source-channel coding} (JSCC) architecture to support efficient speech or multi-media transmission \cite{kurka2020deepjscc,huang2024d}. Specifically, two deep-neural-network models deployed separate at a transmitter and a receiver are jointly trained to not only compress contents (e.g., speech signals or images) into essential features for efficient transmission but also cope  with channel noise \cite{weng2023deep, xie2021deep}. As a result,  the content to be accurately reproduced at the receiving end.  Furthermore, to operate in the case without task knowledge at the transmitter, the JSCC architecture can be enhanced with task  domain transfer at the receiver \cite{zhang2022deep}. As SemCom is still at its nascent stage, existing techniques are too simple to support robotic applications where  robots continuously interact with the physical environment or human beings~\cite{weng2021semantic}. For robotic edge AI, SemCom requires incorporating rules and common senses of the physical world and human language logic into communication designs. 

% One solution is to represent such information using KGs, each consisting   of a set of semantic triples in the form of “entity-relation-entity” \cite{vinciarelli2015open}. Initial attempts have been made on incorporating KGs into 6G networks and designing enabling techniques~\cite{strinati20216g,wang2022performance}. In particular, a so called  \emph{attention proximal policy optimization} algorithm has been proposed to evaluate the importance of triples in a KG, which is useful for efficient transmission and resource allocation~\cite{wang2022performance}. Prior  studies mainly concern  human knowledge representations. There exist no concrete  systematic framework for enabling robotic edge AI. Among others, to realize SemCom for such systems faces the challenge of ultra-low-latency data intensive SemCom, which motivates this work. The \emph{ultra-reliable low-latency communication} (URLLC) in 5G relies on short packets for  sending mission-critical, low-rate messages (e.g., commands and instructions)~\cite{RN283}. However, such a technology cannot support 6G robotic applications that typically involve robots streaming high-rate multi-modal sensing data (e.g., cameras and LIDAR) to servers  and are also latency sensitive, e.g., manufacturing, elderly care, swarm robotics, and disaster relief. 

One solution is to represent such information using KGs, each consisting   of a set of semantic triples in the form of “entity-relation-entity” \cite{vinciarelli2015open}. Initial attempts have been made on incorporating KGs into 6G networks and designing enabling techniques~\cite{strinati20216g,wang2022performance}. In particular, a so called  \emph{attention proximal policy optimization} algorithm has been proposed to evaluate the importance of triples in a KG, which is useful for efficient transmission and resource allocation~\cite{wang2022performance}. 
Prior  studies mainly concern  human knowledge representations. There exist no concrete  systematic framework for enabling robotic edge AI. Among others, to realize SemCom for such systems faces the challenge of ultra-low-latency data intensive SemCom, which motivates this work. The \emph{ultra-reliable low-latency communication} (URLLC) in 5G relies on short packets for  sending mission-critical, low-rate messages (e.g., commands and instructions)~\cite{RN283}.
Diverse approaches are proposed to characterize transmission reliability by considering factors such as packet length, coding rate, and channel parameters \cite{RN275,Liva-TCOM-2019}, with the aim of minimizing packet decoding errors \cite{Schmeink-JSAC-2018,Nallanathan-TWC-2020}. However, the primary concern of robotic SemCom is often the task-oriented metrics like inference accuracy. Existing URLLC schemes, which prioritize packet-decoding performance, overlook the critical aspect of classification accuracy. Moreover, these schemes cannot support 6G robotic applications that typically involve robots streaming high-rate multi-modal sensing data (e.g., cameras and LIDAR) to servers and are also latency sensitive, e.g., manufacturing, elderly care, swarm robotics, and disaster relief. To fully distill the potential of 6G robotic SemCom, a new paradigm integrating communication and AI should be explored.

In this work, we consider a robotic edge-AI system where a server acts as a ``remote brain'' of a robot. 
The system's objective is to find on a KG a \emph{knowledge path} (KP) that represents a feasible sequence of actions the robot can take to accomplish a given task \cite{saxena2014robobrain}.
This requires the robot to upload its observations for the server to recognize objects in the robot's environment.
To the best of our knowledge, this work represents the first attempt to establish a framework of knowledge-based robotic SemCom to provide a goal-oriented air-interface for the system.
Its main components are summarized as follows.
\begin{itemize}
    \item \textbf{Robotic SemCom protocol.} The protocol is proposed to achieve the aforementioned system objective. To this end,  the  design incorporates the following key system operations.  
    1) Given a robotic task, an existing path-searching algorithm is applied to identify all possible  KPs for accomplishing the task. 
    2) Then, based on a semantic similarity metric, semantic matching establishes a mapping between KP triples and class labels of  the server classifier for object recognition. 
    3) The robot streams the observations from exploration of the surrounding environment and transmits the extracted features to the edge server. The proposed ultra-low latency feature transmission scheme for the purpose  is elaborated in the sequel. 
    4) The preceding process is repeated  until the edge server can confidently choose one KP that is feasible for the robot to execute the given  task in its environment. 
    
    \item \textbf{Ultra-low-latency robotic SemCom.} To tackle the aforementioned  challenge of ultra-low-latency data-intensive SemCom, we propose the mentioned scheme of \emph{ultra-low-latency feature transmission} (ULL-FT). Its novelty  lies in exploiting the server-classifier's robustness to counteract channel  errors such that a high coding rate can be supported. This avoids the rate loss in the traditional URLLC to ensure high transmission reliability. The performance of the proposed scheme is quantified based on the widely adopted  GM model for sensing data distribution. As a result, we derive a relation  between the classification margin, which measures the classifier's robustness,  and the link \emph{bit error probability} (BEP), denoted as $P_b$. Based on the relation, the classification accuracy is  shown to follow the scaling law of $1-\mathcal{O}(\sqrt{P_b})$. 
    % In the case where the classification margin is insufficient for counteracting channel distortion, we consider retransmission to rein in the distortion and derive the minimum number of retransmissions for achieving a given classification accuracy for given a given classification margin. 
    
    \item \textbf{Classification margin enhancement.} Consider ULL-FT in the case where the classification margin is insufficient for counteracting channel distortion. We explore two approaches to enhance the margin. First, we consider retransmission to rein in the distortion at the cost of increased latency and derive the minimum number of retransmissions for achieving a given classification accuracy. Alternatively, to enlarge the classification margin, the other via data-cluster-variance reduction, we consider multi-view classification that observations of each object. In view of the lack of theoretic analysis on multi-view classification accuracy, we fill the void and derive the result that the accuracy scales with the number of views $m$ at an exponential rate of $1-\mathcal{O}(e^{-m})$. 
    
    \item \textbf{Experimental results.} Experiments are conducted to validate our designs using both the GM data distribution with linear classification and real datasets with \emph{deep neural network} (DNN) classifier models. The results demonstrate that ULL-FT can achieve significant communication latency reduction as opposed to the traditional reliable transmission while providing a guarantee on accurate feasible KP identification. 
\end{itemize}

The rest of this paper is organized as follows. Section II introduces the models and metrics. Then the robotic SemCom protocol is described  in Section III. The schemes of ULL-FT and multi-view robotic SemCom are presented  and analyzed in Section IV and  Section V, respectively. Experimental results are provided in Section VI.

\section{Models and Metrics}\label{Sec: models and metrics}
\begin{figure}[t!]
\centering
\includegraphics[width=0.99\columnwidth]{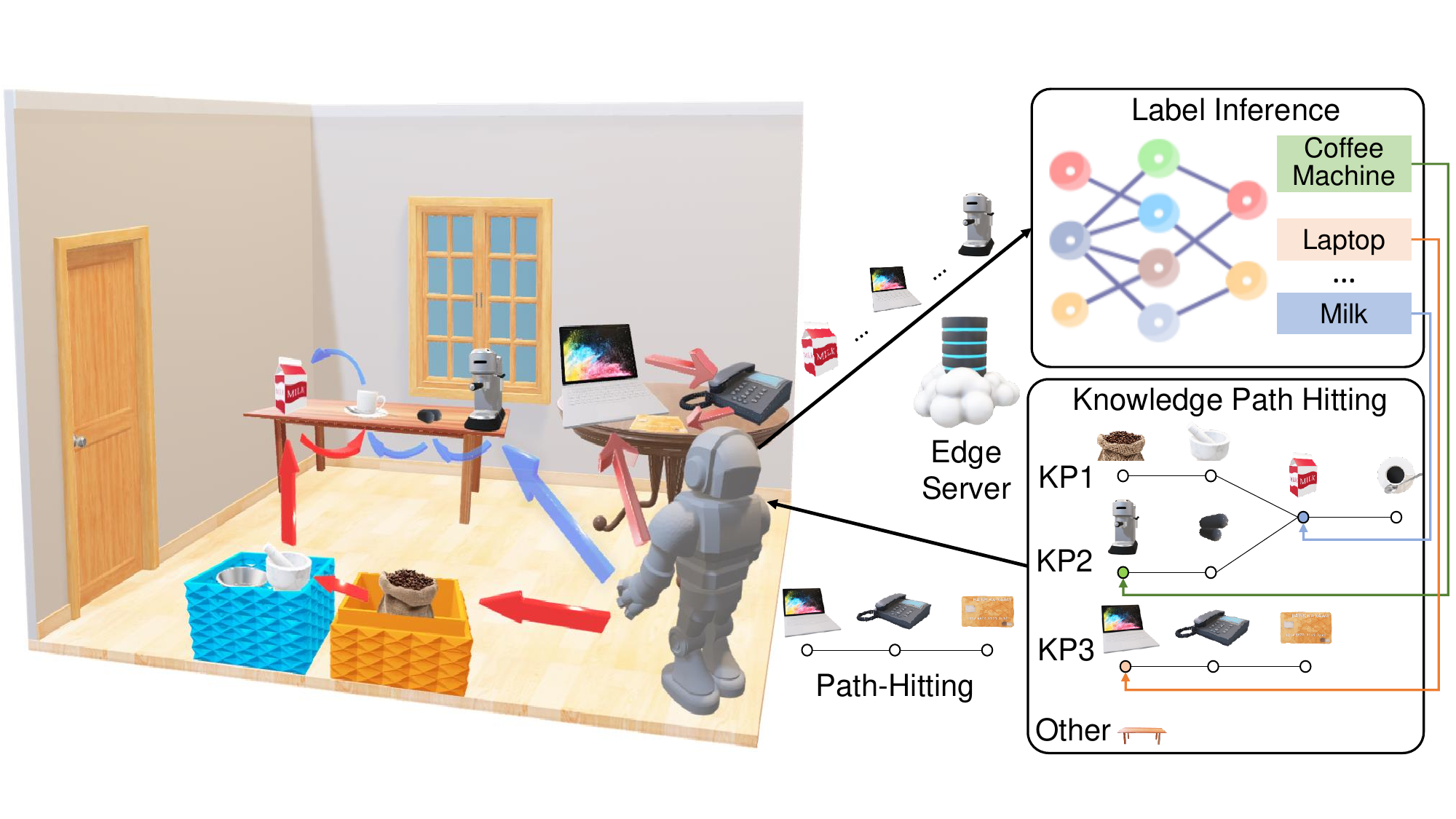}
\caption{Knowledge-based robotic SemCom system. The task of a robot, comprising a sequence of objects and actions, corresponds to paths within a KG. An edge server, equipped with KG, serves as a ``remote brain'' that identifies feasible KPs by analyzing the observations uploaded by the robot.}
\label{fig: system model}
\end{figure}

We consider a robotic-edge-AI system comprising a server-robot pair as illustrated  in Fig. \ref{fig: system model}. Its detailed operations are described in Section \ref{Sec: Protocols}. Relevant models and metrics are discussed in sub-sections. 

\subsection{KP Finding}
A large-scale KG is operated at the  edge server (see e.g., RoboBrain by Standford~\cite{saxena2014robobrain}). It is a directed graph denoted as $\mathcal{G}=(\mathcal{V},\mathcal{A})$, where $\mathcal{V}$ is the vertex set whose elements are object descriptions and $\mathcal{A}$ the arc set containing the actions in terms of the ordered pairs of objects. 

Based on human commands, the robot is required to explore the environment and then complete the task that is equivalent to a path on the KG. 
There exists a rich literature of path-finding algorithms~\cite{ji2021survey}, such as DeepPath~\cite{xiong2017deeppath}, MINERVA~\cite{das2017go} and M-Walk~\cite{shen2018m}. Using one of the algorithms, we assume that the task-relevant KPs have been generated for a given task. Considering a given robotic task, there exist multiple possible  paths for executing the task, each consisting of corresponding objects and actions. Let the $i$-th KP be denoted as $\mathcal{P}_i=\nu_{i_0}\stackrel{\alpha_{i_1}}{\longrightarrow}\nu_{i_1}\stackrel{\alpha_{i_2}}{\longrightarrow}\cdots\stackrel{\alpha_{i_u}}{\longrightarrow}\nu_{i_u}$, where $\{\nu_{i_j}\}_{j=0}^u\subseteq\mathcal{V}$, $\{\alpha_{i_j}\}_{j=1}^u\subseteq\mathcal{A}$, and the path length is $u$.
% \hl{xxx: Write a path as a sequence of vertices (objects) connected by directed arrows  with action specified as overtext.}
Specifically, the set of objects  the $i$-th KP is denoted as $\mathcal{T}_i\subseteq \mathcal{V}$, and the total task-relevant objects are the set $\mathcal{T}=\bigcup_{i}\mathcal{T}_i$. 
The system objective is to choose the feasible KP by identifying relevant objects in the robot's environment, i.e., the identification  of a feasible  KP by objects recognition. 
To this end, the robot explores the environment, captures images of objects in its surrounding environment,  and streams the extracted feature vectors to the server for inference. 
The surrounding objects observed by the robot constitute a set $\mathcal{K}=\{1, 2,\dots, K\}$. 
By classifying the uploaded feature vectors, the objects belonging to  $k\in\mathcal{K}\cap\mathcal{T}$ are deemed as ``relevant'' or otherwise ``irrelevant". 

{\bf Example:} 
To bring a cup of coffee to a user, there are different task-relevant KPs. In one KP, the robot searches for a coffee machine, places a coffee capsule, finds a coffee mug under the machine's nozzle, presses the start button, adds milk, and finally brings the coffee to the user. In another KP, the robot orders coffee from the Starbucks App, travels to the nearest delivery box, picks up the coffee using a received security code, and delivers it to the user. While completing the task, the robot may encounter irrelevant objects like a sofa, TV, or neighbors, but the focus remains on the task-relevant objects such as the coffee machine, coffee capsule, coffee mug, milk, and the user.

\subsection{Robot Observation Model}
\subsubsection{Data Distribution} 
We consider real datasets in experiments and the widely used synthetic data model of GM distribution in analysis for tractability, which is described as follows \cite{lan2022progressive}. 
Each local observation, e.g., a captured image, is characterized by an $N$-dimensional vector and  associated with one of the $L$ classes. The $L$ classes have uniform prior probabilities. 
The server computes  a reduced $D$-dimensional feature space with $D \leq N$ to facilitate feature extraction by using e.g., \textit{principal components analysis} (PCA) or autoencoders. 
Thereby, $\mathbf{x}_{k}\in \mathbb{R}^D$ is generated at the robot to report the observation of the $k$-th object. 
Then  $\mathbf{x}_{k}$ follows a GM distribution in the feature space. 
Each class, say class-$\ell$, is represented by a multivariate Gaussian distribution $\mathcal{N}(\bmu_{\ell}, \mathbf{C}_{\ell})$ with $\bmu_{\ell}\in\mathbb{R}^D$ and $\mathbf{C}_{\ell}\in\mathbb{R}^{D\times D}$ denoting the centroid and the covariance matrix, respectively. 
For simplicity, all covariance matrices are assumed identical: $\mathbf{C}_{\ell} = \mathbf{C}, \forall\ell$, where the covariance matrix $\mathbf{C}$ is  diagonalized  through PCA. 
This is a common assumption made in the literature of statistical learning \cite{hastie2009elements,bishop2006pattern}.
It follows that the \emph{probability density function} (PDF) of $\mathbf{x}_{k}$ can be written as
\begin{equation}\label{GM}
    f(\mathbf{x}_{k}) = \frac{1}{L}\sum_{\ell=1}^{L}\mathcal{N}(\mathbf{x}_{k} \mid \bmu_{\ell}, \mathbf{C}),
\end{equation}
where $\mathcal{N}(\mathbf{x}_{k} \mid \bmu_{\ell}, \mathbf{C})$ denotes the Gaussian PDF with mean $\bmu_{\ell}$ and covariance matrix $\mathbf{C}$. 

\subsubsection{Multi-view Data}
In Section \ref{Sec: Multi-view SemCom}, we consider the the scenario where the robot sequentially captures multiple observations of each object over multiple epochs.  
For a single object in its $i$th epoch, say object-$k$, the robot extracts and transmits the feature vector $\mathbf{x}_k(i)$ to the edge server.
Given a fixed number of epochs per object denoted as $m$, the edge server receives $m$ feature vectors for  each object. 
For multi-view observations, the individual feature vectors are aggregated into a single feature vector, $\overline{\mathbf{x}}_{k}$,  by average pooling before feeding it into a pre-trained multi-view classifier to infer the object  label~\cite{liu2023over}:  
% \begin{equation}
    $\overline{\mathbf{x}}_{k}{(m)} = \frac{1}{m}\sum_{i=1}^m\mathbf{x}_k(i).$
% \end{equation}

\subsection{Communication Model}
Consider the uplink transmission of an arbitrary  feature vector, $\mathbf{x}\in\mathbb{R}^{D}$. 
Each of its element is represented by a signed integer of $n$ bits. Specifically, the decimal value of  the $d$-th feature/element $x_d$ is represented  as 
\begin{equation}
    x_d=\sum_{i=0}^{n-2}b_i2^i-b_{n-1}2^{n-1},
\end{equation}
where $b_i\in\{0,1\}$ is the binary value at the $i$-th position. 
%We consider the following feature transmission scheme.
%The robot transmits the feature vector without sufficient channel coding for reliability. 
During transmission, the robot is allocated a fixed bandwidth. For the ultra-low-latency transmission in Section \ref{Sec: Ultra-Low-Latency Feature Transmission}, we allow a high coding rate such that the uplink has a significant BEP, denoted as $P_b$.
% \hl{xxx: check the manusript to change uncoded to high coding rate.}
% The transmission rate of the robot is fixed as $R$, and thus the latency is fixed as $T=nD/R$ seconds per feature vector.
\begin{comment}
For illustration, in the case of the \emph{binary phase shift keying} (BPSK) modulation and uncoded transmission, the BEP, denoted as $P_b$, has the relation with respect to the SNR as follows:
% \begin{equation}
    $P_b = Q\left(\sqrt{2\rho}\right)$,
% \end{equation}
where $Q(\cdot)$ is the Q-function and defined as $Q(x)=\frac{1}{2\pi}\int_{x}^{\infty}\exp{(-\frac{t^2}{2})}dt$. 
\end{comment}
Since each bit has the probability of  $P_b$ to flip, the received distorted  value $\hat{x}_d$ is given as
\begin{equation}\label{eqn: received feature}
    \hat{x}_d=\sum_{i=0}^{n-2}[a_i(-1)^{b_i}+b_i]2^i-[a_{n-1}(-1)^{b_{n-1}}+b_{n-1}]2^{n-1},
\end{equation}
where the binary indicators $\{a_i\}$ are i.i.d. Bernoulli random variables, i.e., $a_i\sim \text{Bernoulli}(P_b)$. 

\subsection{Classifier Models and Metrics}
\subsubsection{Linear Classifier} The model is adopted in analysis for tractability \cite{hastie2009elements}. 
For a pair of clusters, e.g., cluster-$\ell$ and cluster-$\ell'$, the decision boundary is a hyperplane between them defined as $\mathcal{H}(\mathbf{w},b)=\{\mathbf{x}:\mathbf{w}^T\mathbf{x}+b=0\}$.
% \hl{xxx: need write using w and b}. 
The label of a feature vector $\mathbf{x}$ can be determined by computing its output score defined as follows: 
\begin{equation}\label{eqn: output score}
    s(\mathbf{x})=\mathbf{w}^T\mathbf{x}+b.
\end{equation}
The feature vector is labeled as one class, say class-$
\ell$, if $s(\mathbf{x}) > 0$ or the other class, say class-$\ell'$, if $s(\mathbf{x}) < 0$. 
% \hl{xxx: Describe the case of multiple classes (multiple hyperplanes) and how it can be decomposed as binary classification.}
A general multi-class classifier, e.g., an $L$-class classifier, can be implemented using the method of  \emph{one-versus-one}, which decomposes the classifier into $J=L(L-1)/2$ binary classifiers \cite{bishop2006pattern}. Considering a feature vector $\mathbf{x}$ and a pair of cluster centroids,  each binary operation  generates the  distances from each centroid to the projection of  $\mathbf{s}$ onto the line connecting the centroid. Then upon completion of $J$ binary classifications, $\mathbf{x}$ is assigned the label of the nearest cluster. 
Three  useful metrics are discussed  as follows.
\begin{itemize}
    \item \textit{Mahalanobis Distance:}
    Mahalanobis distance is a statistical measure used to quantify the similarity between two points in a multivariate dataset. Mathematically, the distance between the feature vector $\mathbf{x}_{k}$ and the centroid $\bmu_{\ell}$ in the feature space, with covariance matrix $\mathbf{C}$, is defined as~\cite{de2000mahalanobis}
    \begin{equation}\label{eqn: HSMD}
        d_{\mathbf{C}}(\mathbf{x}_k,\bmu_{\ell}) = \sqrt{(\mathbf{x}_k-\bmu_{\ell})^T\mathbf{C}^{-1}(\mathbf{x}_k-\bmu_{\ell})}.
        % =\sqrt{\sum\nolimits_{d=1}^D {(x_{k,d} -\mu_{\ell,d})^2}/{C_{d}}}.
    \end{equation}
    Then, the problem of linear classification of object-$k$ can be reduced to minimizing the Mahalanobis distance: $\hat{\ell} = \arg\min\limits_{\ell} $ $d_{\mathbf{C}}(\mathbf{x}_k,\bmu_{\ell})$.
    % \item \textit{Inference Uncertainty:}
    % In the machine learning literature, inference uncertainty is typically measured by the entropy of posteriors \cite{hastie2009elements}.
    % Considering the GMM model, the entropy,  denoted as $H_{k}$ for the classification of the $k$-th object, is given as 
    % \begin{equation}\label{uncertainty}
    %     H_{k} = -\sum_{\ell = 1}^{L}\Pr(\ell|\overline{ \mathbf{x}}_{k}) \log\Pr(\ell|\overline{ \mathbf{x}}_{k}).
    % \end{equation}
    \item \textit{Discrimination Gain:}
    The discrimination gain quantifies the differentiability   between two classes in the feature space. 
    For a given class pair, e.g., class-$\ell$ and class-$\ell'$, the pairwise discriminant gain can be written in terms of  Mahalanobis distance as~\cite{lan2022progressive}
    \begin{equation}\label{eqn: discriminant gain}
        g{(\ell,\ell')} = d_{\mathbf{C}}(\bmu_{\ell},\bmu_{\ell'}).
    \end{equation}
    \item \textit{Correct Classification Accuracy:}
    We consider binary classification and assume that the transmitted feature vector can be perfectly classified, i.e., the two clusters are separable with high probability. (Note: The assumption is relaxed in Section \ref{Sec: Multi-view SemCom}.) 
    A pair of transmitted and received feature vectors should be at the same side of the decision hyperplane of the classifier such that they have the same classified labels. 
    Given the above assumption, this event is referred to as \emph{noisy feature alignment} and its probability gives the correct classification probability~\cite{liu2020wireless}. 
    Given a transmitted feature vector, the received  feature  error vector, distorted by the channel is denoted as $\hat{\mathbf{x}}$. 
    Then the noisy feature alignment event is specified as the set of events that occur with $s(\mathbf{x})s(\hat{\mathbf{x}})>0$, and the correct classification probability is defined as
    $\Pr\left(s(\mathbf{x})s(\hat{\mathbf{x}})>0\right)$. 
\end{itemize}
\subsubsection{CNN Classifier}
In experiments, we also consider the practical CNN classifier model to validate insights from analysis. 
For CNN classifiers, the architecture comprises multiple \emph{convolutional} (CONV) layers followed by multiple \emph{fully-connected} (FC) layers. 
To implement split inference, the classifier is partitioned into the robot and server sub-models, represented as functions $f_{\sf rob}(\cdot)$ and $f_{\sf ser}(\cdot)$, respectively. 
Let $N_c$ denote the number of CONV filters in the final layer of $f_{\sf rob}(\cdot)$, and each filter generates a feature map with dimensions $L_h$ and $L_w$.
The set of all extracted feature maps from an input image is represented by the tensor $\mathbf{X}\in\mathbb{R}^{N_c\times L_h\times L_w}$. The performance metric is presented as follows.
\begin{itemize}
    \item \textit{Top-1 Accuracy:}
    It is a widely-used performance metric for evaluating classification models with multiple classes~\cite{petersen2022differentiable}. This metric represents the proportion of received samples where the class with the highest predicted probability matches the true class or ground truth label. In essence, Top-1 accuracy measures the percentage of samples where the prediction of the CNN classifier is accurate.
\end{itemize}

% \hl{xxx: For the following part, desribe first single view used in first part of paper. And then extend it to multi-view with minimum repetition with single view equations. }
\subsubsection{Multi-view Classification}
For linear classifiers, the feature vector of the $k$-th object extracted at the $i$-th view is denoted as $\mathbf{x}(i)$. To reduce data variance, multi-view operations employ averaging, resulting in $\overline{\mathbf{x}}_k(m)=\frac{1}{m}\sum_{i=1}^m \mathbf{x}_k(i)$. Subsequently, a relationship is established between the aggregated feature vector $\overline{\hat{\mathbf{x}}}_k$ and the received feature vectors $\{\hat{\mathbf{x}}(i)\}$ at the edge server up to the $m$-th view through average pooling: $\overline{\hat{\mathbf{x}}}_k(m)=\frac{1}{m}\sum_{i=1}^m \hat{\mathbf{x}}_k(i)$. The noisy feature alignment event is characterized by $s(\overline{\mathbf{x}}_k)s(\overline{\hat{\mathbf{x}}}_k)>0$, and the performance is measured by the correct classification probability $\Pr(s(\overline{\mathbf{x}}_k)s(\overline{\hat{\mathbf{x}}}_k)>0\mid m)$.
For CNN classifier, the multi-view dataset comprises various objects, each with multiple views. The average pooling process remains consistent, provided that the feature vector $\mathbf{x}$ is replaced by the feature map $\mathbf{X}$. The performance metric remains the Top-1 accuracy by classifying the average pooled feature maps.

\section{Robotic SemCom Protocol}\label{Sec: Protocols}
\begin{figure}[t!]
\centering
\includegraphics[width=0.82\columnwidth]{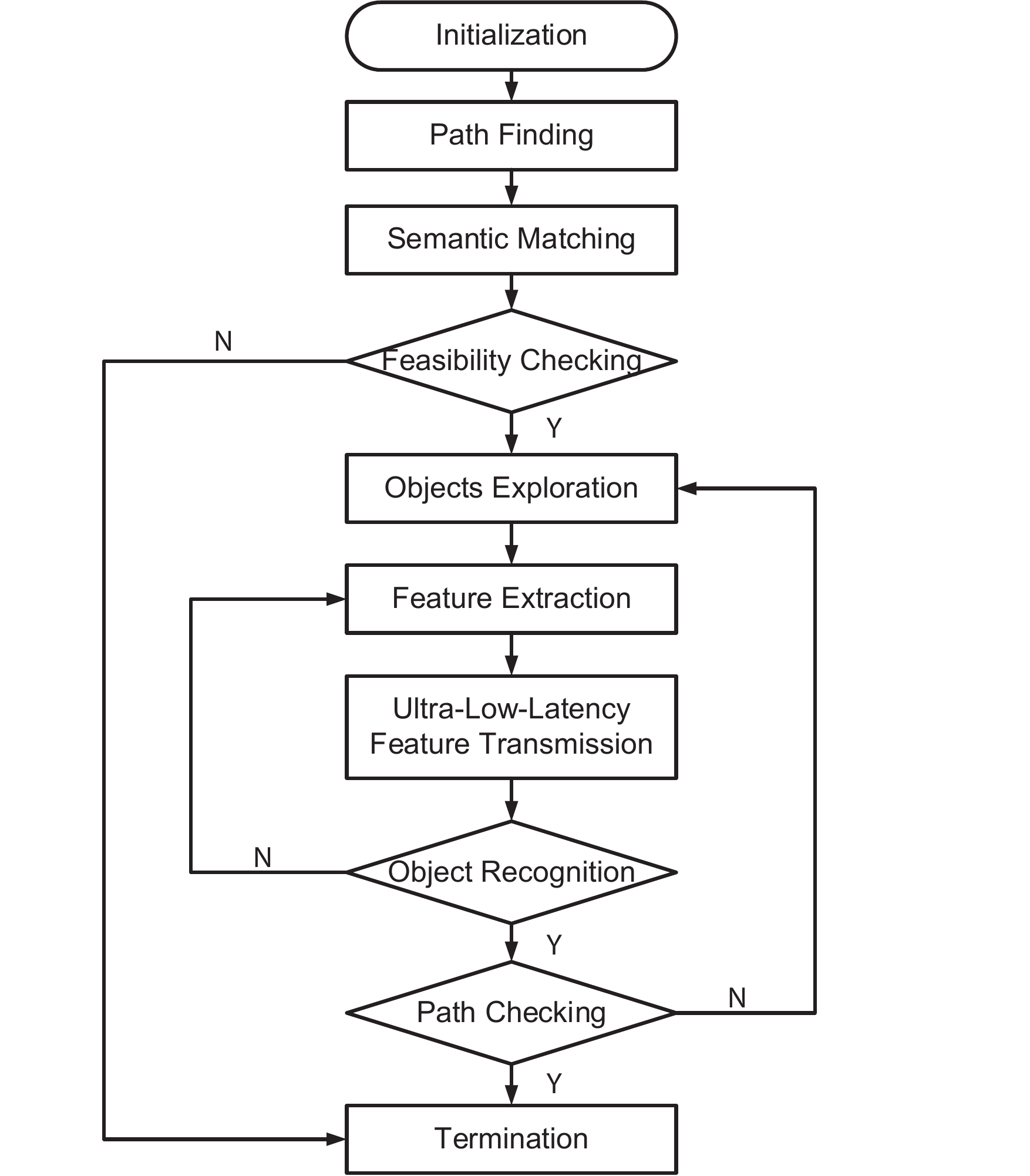}
\caption{The workflow chart of the Robotic SemCom protocol.}
\label{fig: flow chart of protocol}
\end{figure}

The objective of the robotic SemCom protocol is to use a KG to identify a feasible sequence of  actions for the robot to take for accomplishing its given task. The protocol is summarized in the flowchart in Fig.~\ref{fig: flow chart of protocol} and its key steps are elaborated as follows.
\begin{enumerate}
    \item \emph{KP Finding:} At the robot, human commands are translated into a task using sensing and natural language  processing, which is sent to the server as a service request \cite{saxena2014robobrain}. Then 
    using an existing path-finding algorithms, the server computes  a set of task-relevant  KPs on its  large-scale KG, $\mathcal{G}=(\mathcal{V},\mathcal{A})$, targeting the requested task  (see e.g.,~\cite{ji2021survey,xiong2017deeppath,das2017go,shen2018m}). 
    % \hl{xxx: Please check to replace "available" or "possible KPs" with "task-relevant KPs" for consistency. }
    The task-relevant objects, referring to  the union of the object sets of task-relevant KPs, are aggregated into  the set $\mathcal{T}\subseteq\mathcal{V}$.
    \item \emph{Semantic Matching:} 
    % \hl{Fix consistency issue, it was called "semantic mapping" in introduciton.} 
    All possible output labels from the classifier are denoted as the set $\mathcal{L}=\{1,\cdots,L\}$. 
    The nodes in the KG are stored in terms of objects' name that may not match the classes' labels even if they are semantically equivalent, e.g., ``table'' and ``desk''.
    Semantic matching aims to address the issue by  mapping objects in $\mathcal{T}$ to the labels in $\mathcal{L}$. 
    The key operation is to embed the words of labels as well as description of KG nodes to the semantic space using a pre-trained model a neural network model, denoted as $\psi(\cdot)$ (see e.g., \texttt{word2vec} by Google). 
    The  embedding association  function measures the distance $d_{\sf Sem}(\cdot,\cdot)$ (e.g., Euclidean or cosine distance) between two embedding vectors  in the semantic space. 
    Then for each label $\ell\in\mathcal{L}$, we assign its relevance score as follows:
    \begin{equation}
        \mathbb{I}(\ell)=\begin{cases}
            1,& \text{if }\exists j\in\mathcal{T}, \text{ s.t. } d_{\sf Sem}(\psi(\ell),\psi(j))\leq\epsilon,\\
            0,&\text{otherwise}.
        \end{cases}
    \end{equation}
    Here, $\epsilon$ is the pre-determined semantic similarity threshold. The task-relevant classes are aggregated in the set
    $
    % \begin{equation}
        \mathcal{R}=\{\ell\in\mathcal{L}:\mathbb{I}(\ell)=1\}.
    % \end{equation}
    $
    \item \emph{Feasibility Checking:} There exist two possible scenarios that renders  the robotic task infeasible: i) there exits no KP in the task-relevant KP set whose objects can be all recognized by the classifier  model; ii) there exits no KP in the task-relevant KP set whose objects all exist in the robot's environment. 
    The first scenario can be checked  immediately after semantic checking, i.e., the feasibility requires: $\exists i, ~\text{s.t.}~ \mathcal{T}_i\cap\mathcal{R}\neq \emptyset$.
    If the condition holds, the second scenario can only be checked  after the robot explores the environment to find the relevant objects $k\in\mathcal{K}\cap\mathcal{R}$, i.e., the feasibility requires: $\exists i,~\text{s.t.}~\mathcal{K}\cap\mathcal{T}_i\neq\emptyset$.
    % \item \emph{Stop Control:} The accuracy of the feature vector (or the number of features) of a particular image needed for classification to meet the accuracy requirement is object dependent task. Neither the robot nor the server has prior knowledge of this number since each has access to either the object or the model but not both. Online stopping control aims at minimizing the number of transmissions under the said requirement. Its procedure is described as follows. Let $b_i$ indicate the server’s decision on whether the robot should retransmit the feature vector (or transmit features) in slot-$i$ (i.e., $b_i = 1$), or stop the transmission (i.e., $b_i = 0$). If the decision is to transmit, proceed to the next step; otherwise, go to Step 7.
    \item \emph{Ultra-Low-Latency Feature Transmission:} As elaborated in Section~\ref{Sec: Ultra-Low-Latency Feature Transmission}, our technique integrates transmission and classification to support ultra-low-latency feature vector uploading. 
    \begin{itemize}
        \item \emph{(Enhancement) Retransmission or Multi-view Feature Aggregation:} 
        In the case where the classification margin is sufficient for coping with channel distortion, we propose two enhancement solutions: i) retransmission, i.e., the robot transmits the feature vector for multiple times; ii) multi-view feature aggregation, i.e., the robot obtains  multi-view observations of  each local object. Both approaches can enhance the margin and thereby boost the inference accuracy. The details are provided in Section \ref{Subsection: Margin Enhancement}. 
    \end{itemize}  
    \item \emph{Server Inference:} The server feeds each received feature vector (or the average pooled feature vector in the case of multi-view SemCom) into its classification model  to output  an estimated  object  label with an  associated accuracy level. 
    \item \emph{Path Checking:} Upon passing a given accuracy threshold, the object recognized with confidence  undergoes validation using the task-relevant KP set to check its task relevance by inclusion. 
    The protocol is terminated  once a particular KP whose objects are all found to exist in the environment, which is called ``path-hitting''. 
    If no  path-hitting event occurs, the server instructs the robot to continue exploration.
    \item \emph{Termination:} The protocol is terminated if a path-hitting event occurs (a success) or a time limit is reached (a failure). 
    In the former, the identified feasible KP is downloaded onto the robot for  task execution. 
    In the latter, the task is declared to the robot as well as its human user as infeasible. 
    
\end{enumerate}

% \begin{figure}[t!]
% \centering
% \includegraphics[scale=0.50]{figures/BER_vs_MSE_int8.pdf}
% \caption{MSE v.s. BER (int8).}
% \label{fig: MSE vs BER}
% \end{figure}

\section{Ultra-Low-Latency Feature Transmission}\label{Sec: Ultra-Low-Latency Feature Transmission}
\begin{figure}[t!]
    \centering
    \includegraphics[width=1.0\columnwidth]{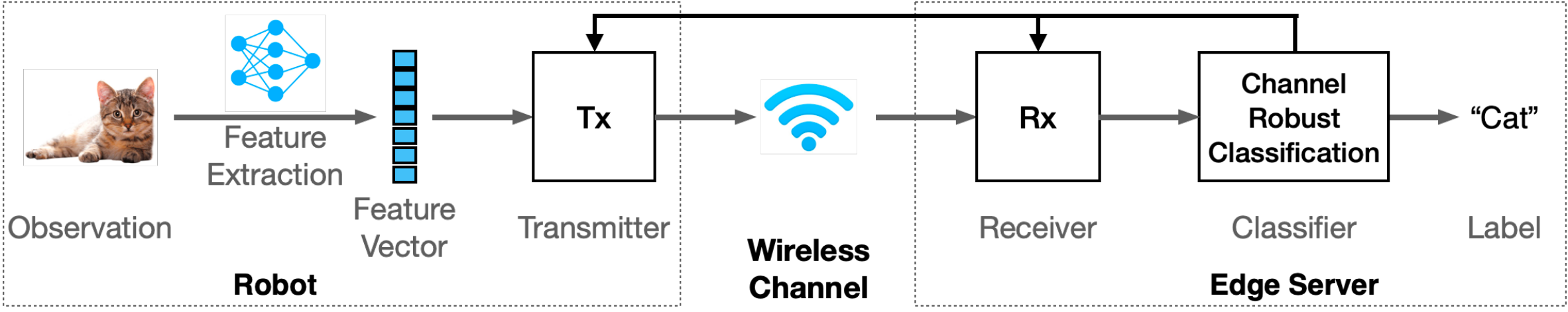}
    \caption{An illustration of the operations in the proposed ULL-FT scheme. The robustness of classification is exploited to accommendation channel distortion.}
    \label{fig: ULL Feature Transmission}
\end{figure}
In this section, we present the ULL-FT scheme as a key component of the proposed robotic SemCom framework. The scheme is first described followed by analysis to provide the design guidelines by quantifying the effects of channel distortion on classification margin and accuracy. 

\subsection{ULL-FT Scheme}
The scheme with key  operations  illustrated in Fig.~\ref{fig: ULL Feature Transmission} is used by  the robot to shorten the duration for  uploading local-observation features  to the server. 
As mentioned, a traditional URLLC design is task-agnostic and   focuses on ensuring link reliability at the cost of data rate. On the contrary, the current scheme features transmission-classification integrated design. Its principle is to delegate part of the task of coping with channel unreliability to the classifier by exploiting its robustness. 
As a result, a high coding rate or even uncoded transmission  can be supported to provide a solution for ultra-low-latency data intensive SemCom.
For the proposed scheme to be effective, the most important is to regulate the channel-distortion level to be within the limit of classifier robustness so that the end-to-end performance degradation is negligible. Then, to design ULL-FT requires the understanding of the relationship between channel distortion, classification robustness, and classification accuracy. An overview of the relationship is provided below.
\begin{figure}[t!]
    \centering
    \includegraphics[width=0.5\columnwidth]{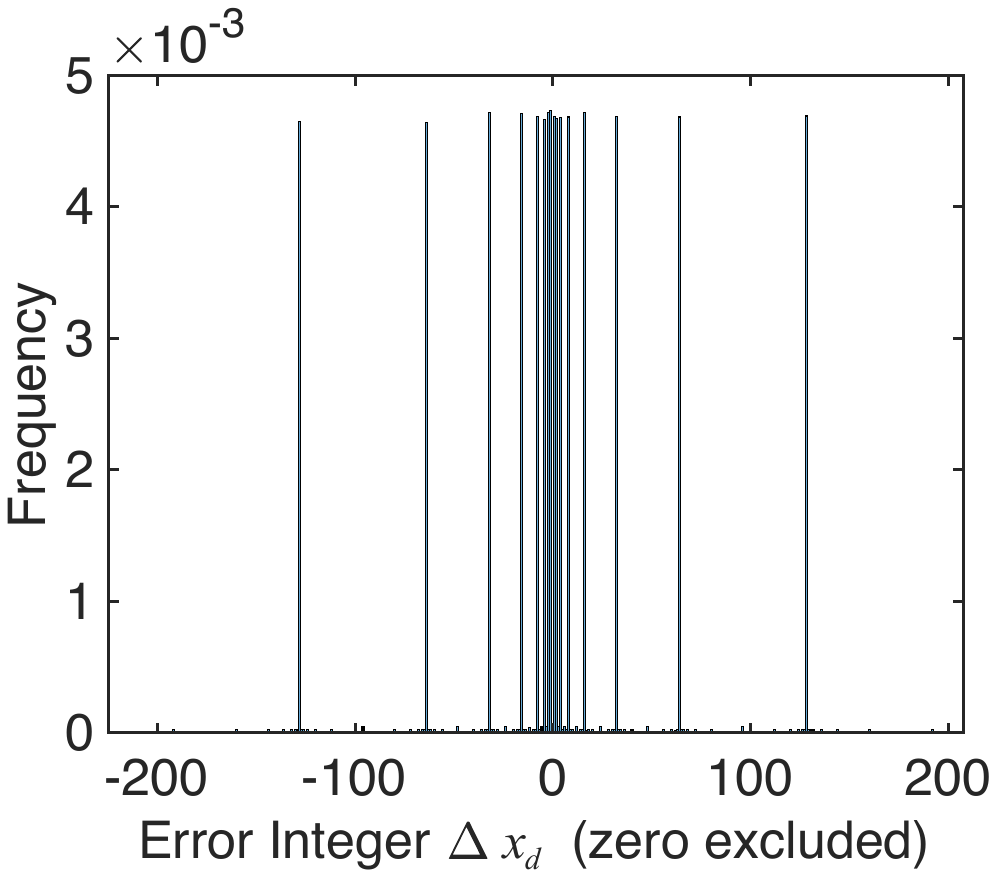}
    \caption{Histogram of feature error $\Delta x_d$, i.e., the $d$-th dimension of feature vector $\mathbf{x}$, with zero value ($\Delta x_d=0$) removed. The BEP is set as 0.01.}
    \label{fig: histograms}
\end{figure}
\subsubsection{Channel Distortion}
The channel distortion of individual features are characterized as follows. Given the preceding transmission scheme, a feature vector is transmitted at a fixed rate without coding or with weak coding. 
For simplicity, we consider uncoded transmission while the extension to the transmission with coding can be easily obtained by increasing the effective \emph{signal-to-noise ratio} (SNR) by the  coding gain (see e.g., \cite{goldsmith1998adaptive}).
We denote the error of an arbitrary  feature, say $x_d$, as the random variable $\Delta x_d=\hat{x}_d-x_d$, where $\hat{x}_d$ is the received $d$-th feature. 
The histogram of $\Delta x_d$ with zero removed is plotted in Fig.~\ref{fig: histograms} with the BEP set as $P_b=0.01$. 
% \hl{(xxx: Specify Int8? etc.)} 
Given $n$ being 8 with data type being Int8, one can observe most of the errors occur at the following $(2n+1)$ discrete positions:
\begin{equation}
    \Delta x_d\!\in\!\{-2^{n-1}\!,-2^{n-2},\cdots\!,-2^0,0,2^0,\cdots\!,2^{n-2}\!,2^{n-1}\}.
\end{equation}
The observation is consistent the fact that the probability of single bit flip is considerably greater than that of multiple bit flips. 

\begin{figure}[t!]
\centering
\includegraphics[width=1.0\columnwidth]{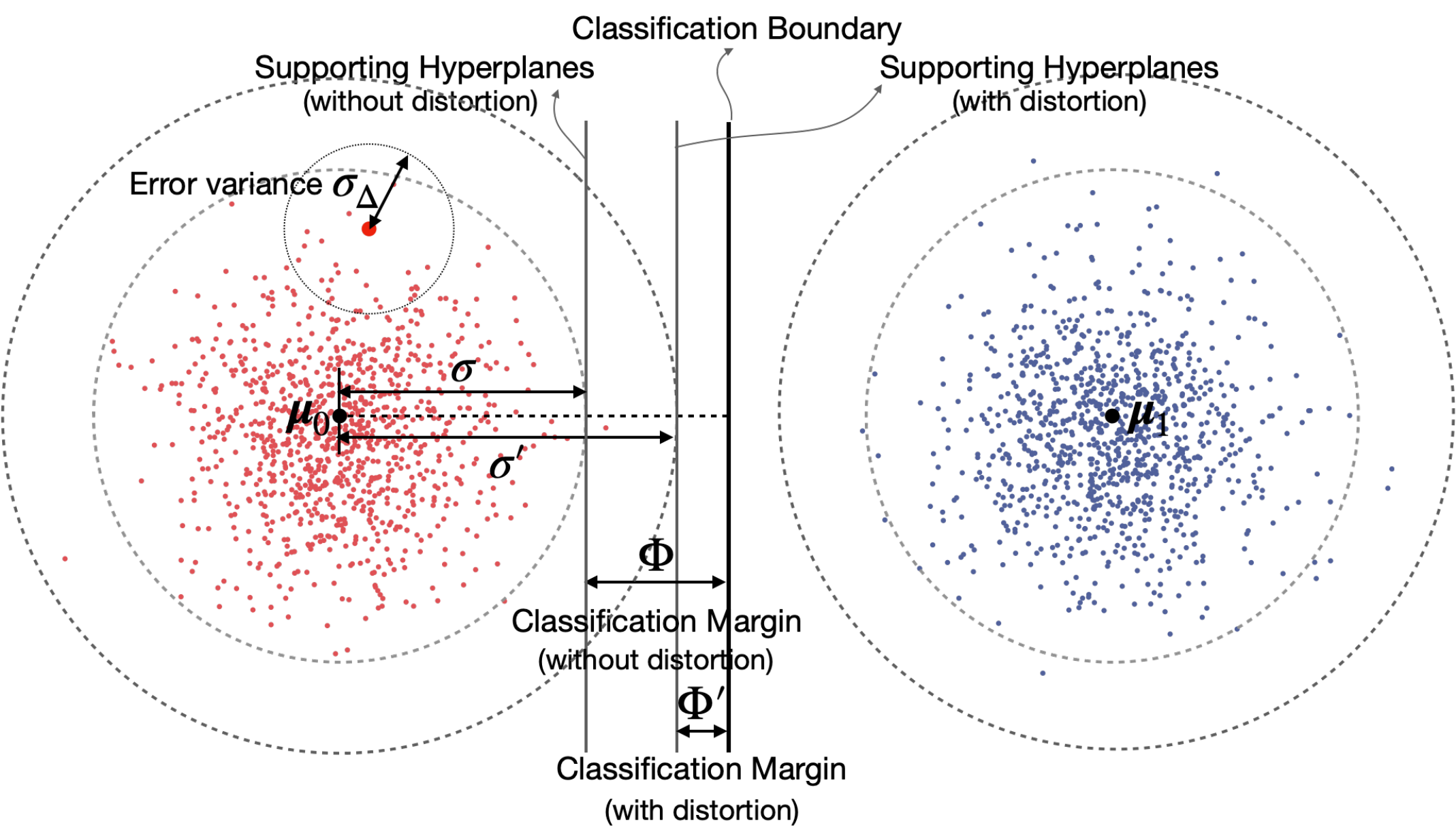}
\caption{The principle of coping with channel distortion using classification margin in GM model.}
\label{fig: margin-based classifier}
\end{figure}

\subsubsection{Classification Robustness}
The robustness of classification is quantified by a metric called \emph{classification margin} \cite{hastie2009elements}, which depends on the data distribution, training time/data, and complexity of the classifier. 
As illustrated in Fig. \ref{fig: margin-based classifier}, the \emph{confidence region} of one cluster refers to a region in the feature space where there is a specific probability level (e.g., 95\% or 99\%) that a feature vector sampled from this cluster will be inside this region.
The decision boundary is a hyperplane that separates the two convex regions and is equidistant from the two.
Following  the literature of statistical learning, the classification margin, denoted by $\Phi$, is  defined as the minimum distance from an arbitrary feature vector in the region to the decision boundary \cite{sokolic2017robust}. 
The mathematical quantification of this metric is provided in Section \ref{Sec: channel distortion and classification margin}. 
\subsubsection{Effectiveness Requirement}
The effect of channel distortion on classification performance is reflected in reduction on classification margin. 
Then, the effectiveness requirement for ULL-FT is to constrain the reduction, which results from feature distortion $\{\Delta x_d\}$, such that the remaining classification margin is non-zero with a high probability. The characterization of the requirement is carried out in the following sub-sections in several steps.
As shown in Section \ref{Sec: channel distortion and classification margin}, $\Phi$ is a monotone decreasing function of BEP $P_b$.
On the other hand, the correct classification probability is derived as a monotone increasing function of $\Phi$ in Section \ref{Sec: channel distortion and classification margin}.
Combining the results established in Section \ref{Sec: channel distortion and classification accuracy} determines the desired effectiveness requirement.
% \begin{figure}[t!]
% \centering
% \includegraphics[width=0.98\columnwidth]{Figures/Margin_Classifier.pdf}\vspace{-2mm}
% \caption{Coping with channel distortion using classification margin.}
% \label{fig: margin-based classifier}\vspace{-3mm}
% \end{figure}
\subsection{Normality of Channel Distortion of Features}
The main difficulty faced in analyzing  the effect of channel distortion on classification margin is that the discrete distribution of feature channel distortion in \eqref{eqn: received feature} does not allow tractability. In the sequel, the difficulty is overcome by proving the normality inherent in feature channel distortion. 
% First, the channel distortion of individual features are characterized as follows. Given the preceding transmission scheme, a feature vector is transmitted at a fixed rate without coding or with weak coding. 
% For simplicity, we consider uncoded transmission  while the extension to the transmission with coding can be easily obtained by increasing the effective SNR by the  coding gain (see e.g., \cite{goldsmith1998adaptive}).
% We denote the error of an arbitrary  feature, say $x_d$, as the random variable $\Delta x_d=\hat{x}_d-x_d$, where $\hat{x}_d$ is the received $d$-th feature. 
% The histogram of $\Delta x_d$ with zero removed is plotted in Fig.~\ref{fig: histograms}(a) with the BEP set as $P_b=0.01$. 
% % \hl{(xxx: Specify Int8? etc.)} 
% Given $n$ being 8 with data type being Int8, one can observe most of the errors occur at the following $(2n+1)$ discrete positions:
% \begin{equation}
%     \Delta x_d \in \{-2^{n-1} ,-2^{n-2},\cdots ,-2^0,0,2^0,\cdots ,2^{n-2},2^{n-1}\}.
% \end{equation}
% The observation is consistent the fact that the probability of single bit flip is considerably greater than that of multiple bit flips. 
As discussed, most of the feature errors occur at the $(2n+1)$ discrete positions. 
Then we approximate error distribution as 
\begin{align}
    &\Pr(\Delta x_d=0)=(1-P_b)^n,\\
    &\Pr(\Delta x_d=-2^{n-1})=\cdots=\Pr(\Delta x_d=2^{n-1})\nonumber\\
    &\qquad\qquad\qquad\qquad~\approx(1-(1-P_b)^n)\times\frac{1}{2n}.
\end{align}
Hence, its first and second moments are $\mathbb{E}[\Delta x_d]=0$, and
\begin{align}
    \mathbb{E}[(\Delta x_d)^2]&\!=\!\left(4^{n\!-\!1}\!\!+4^{n\!-2\!}\!+\cdots+1\right)\!\times\!(1\!-\!(1\!-\!P_b)^n)\!\times\!\frac{1}{2n}\!\times\!2\nonumber\\
    &=\frac{4^n-1}{3n}(1\!-\!(1\!-\!P_b)^n)\approx\frac{4^n-1}{3}P_b,\label{eqn: approx}
\end{align}
where the approximation in \eqref{eqn: approx} is based on that $P_b\ll1$. It follows that the variance of the feature error is given as 
\begin{equation}
    \mathrm{Var}[\Delta x_d]=\mathbb{E}[(\Delta x_d)^2]-\mathbb{E}[\Delta x_d]^2=\frac{4^n-1}{3}P_b.
\end{equation}
For a feature vector $\mathbf{x}$, the error between the received vector $\hat{\mathbf{x}}$ and the original one is a random vector $\Delta\mathbf{x}=\hat{\mathbf{x}}-\mathbf{x}$. 
Assuming i.i.d. elements, its mean is a zero vector and variance is $\mathrm{Var}[\Delta \mathbf{x}]=\mathrm{Var}[\Delta x_d]\mathbf{I}=\frac{4^n-1}{3}P_b\mathbf{I}$.

% \begin{figure}[t!]
%     \centering
%     \subfigure[Histogram of feature error  (with zero removed)]{
%     \includegraphics[width=0.33\columnwidth]{Histogram.pdf}}
%     \subfigure[Histogram of projected error of a feature vector]{
%     \includegraphics[width=0.34\columnwidth]{Projection_Error.pdf}}
%     \caption{(a) Histogram of feature error $\Delta x_d$ with zero value removed; (b) Histogram of the projected feature vector error $\mathbf{w}^T\Delta \mathbf{x}$.}
%     \label{fig: histograms}\vspace{-3mm}
% \end{figure}

Next, based on the preceding results, we show the normality of the projection of the feature error vector onto an arbitrary unity vector: $\mathbf{w}\in\mathbb{R}^{D\times D}$ with $\mathbf{w}^T\mathbf{w}=1$. 
Projecting an abitrary feature vector $\mathbf{x}$ to the direction $\mathbf{w}$ gives the value $\mathbf{w}^T\mathbf{x}$. 
Consequently, the projected error, or the error in this direction, is denoted by the random variable $\mathbf{w}^T\Delta \mathbf{x}$. 
It is obvious that the mean of $\mathbf{w}^T\Delta \mathbf{x}$ is zero; the variance can be derived as follows:
% \begin{equation}
$    \mathrm{Var}[\mathbf{w}^T\Delta \mathbf{x}]=\mathbf{w}^T\mathrm{Var}[\Delta \mathbf{x}]\mathbf{w}=\frac{4^n-1}{3}P_b\mathbf{w}^T\mathbf{w}=\frac{4^n-1}{3}P_b.$
% \end{equation}
\begin{figure}[t!]
    \centering
    \includegraphics[width=0.55\columnwidth]{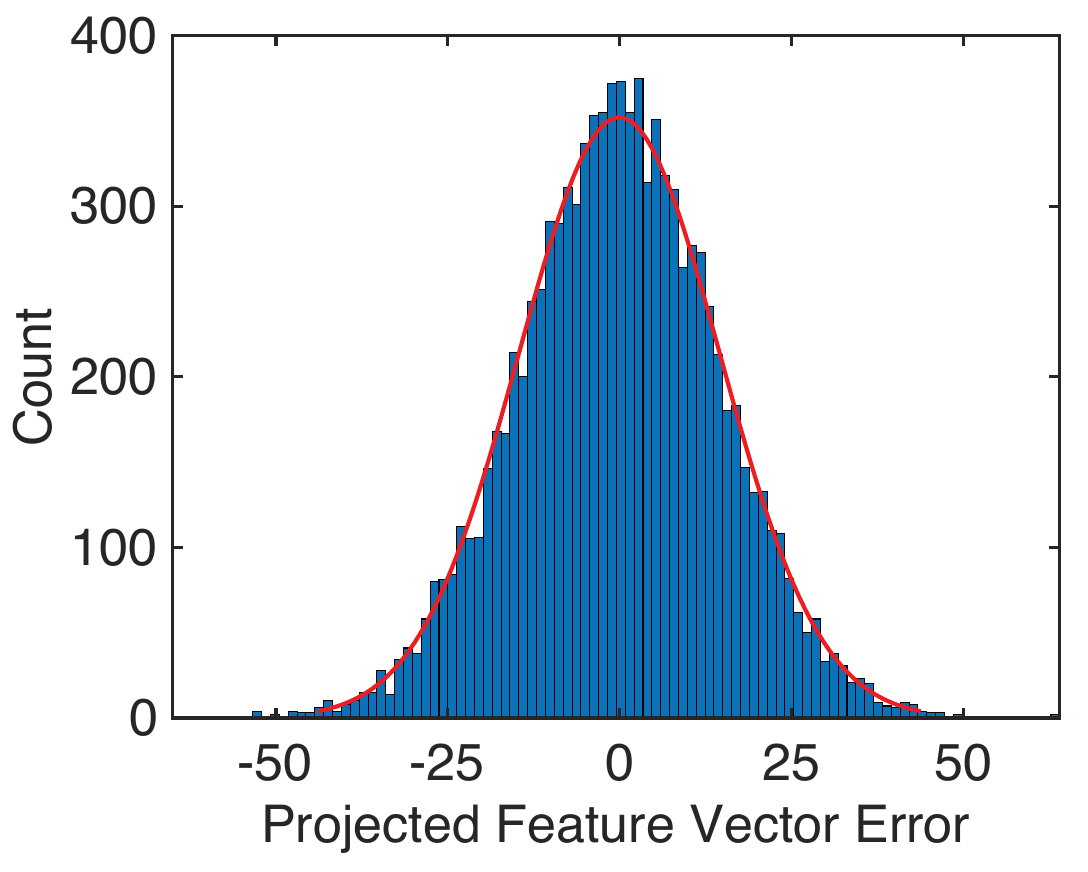}
    \caption{Histogram of the projected feature vector error $\mathbf{w}^T\Delta \mathbf{x}$. The data is fitted using a zero-mean Gaussian distribution.}
    \label{fig: histograms of normalization}
\end{figure}
To show the normality of the projected error,  we introduce a  variant of central limit theorem as follows. 
\begin{Lemma}[\!\!\cite{fisher1992skorohod} Theorem 3.1]
\emph{Let $\{X_d\}$ be a sequence of square integrable i.i.d. random variables with $\mathbb{E}[X]=0$ and $\mathbb{E}[X^2]=1$. Consider a sequence $\{w_d\}$ and the following notations:}
\begin{align}
    S_D&=w_1X_1+w_2X_2+\cdots+w_DX_D,\\
    W_D^2&=w_1^2+w_2^2+\cdots+w_D^2.
\end{align}
\emph{If $W_D^2\to\infty$ and the condition $w_D^2/W_D^2=O(\frac{1}{D})$ holds, then}
\begin{equation}
    \frac{S_D}{W_D}\to Z\quad\text{weakly  as}\quad D\to\infty,
\end{equation}
\emph{where $Z$ is a standard normal random variable.}
\end{Lemma}
Based on the above lemma, we can approximate the distribution of projected error $\mathbf{w}^T\Delta\mathbf{x}$ as a zero-mean Gaussian distribution $\mathcal{N}(0,\sigma_{\Delta}^2)$ with the variance being
\begin{equation}
    \sigma_{\Delta}^2:=\frac{4^n-1}{3}P_b,
\end{equation}
with the approximation is increasing accurate as  the dimensions of the feature vector grow ($D\to\infty$). 
To illustrate the approximation, the histogram of the projected feature error $\mathbf{w}^T\Delta\mathbf{x}$ is shown in Fig.~\ref{fig: histograms of normalization}.

Last, since the approximation  holds for an arbitrary projection direction along $\mathbf{w}$, the error vector $\Delta \mathbf{x}$ follows an isotropic Gaussian distribution, i.e.,
% \begin{equation}
 $   \Delta \mathbf{x}\sim\mathcal{N}(\mathbf{0},\sigma_{\Delta}^2\mathbf{I}).$
% \end{equation}
Without loss of generality, we consider the $k$-th object's feature vector $\mathbf{x}_k$ is sampled from the cluster with centroid $\bmu_{\ell}$. 
The received noisy feature vector is the sum of two multivariate Gaussian vectors: $\hat{\mathbf{x}}_k=\mathbf{x}_k+\Delta\mathbf{x}$, where $\mathbf{x}_k$ follows $\mathcal{N}(\bmu_{\ell},\mathbf{C})$ and $\Delta\mathbf{x}$ follows $\mathcal{N}(\mathbf{0},\sigma_{\Delta}^2\mathbf{I})$. 
Thus, we can conclude that the received feature vector $\hat{\mathbf{x}}_k$ follows the multivariate Gaussian distribution as follows:
\begin{equation}\label{eqn: hat x distribution}
    \hat{\mathbf{x}}_k\mid \ell\sim\mathcal{N}(\bmu_{\ell},\mathbf{C}+\sigma_{\Delta}^2\mathbf{I}).
\end{equation}
Hence, the PDF of the received feature vector of an object, e.g., object-$k$, is GM distribution:
\begin{equation}\label{eqn: PDF of noisy feature vector}
    f(\hat{\mathbf{x}}_{k}) = \frac{1}{L}\sum_{\ell=1}^{L}\mathcal{N}(\hat{\mathbf{x}}_{k}\mid \bmu_{\ell}, \mathbf{C}+\sigma_{\Delta}^2\mathbf{I}).
\end{equation}

\subsection{Channel Distortion and Classification Margin}\label{Sec: channel distortion and classification margin}
Given the normality of feature projection error, we can readily derive the relation between channel distortion and classification margin. As mentioned, linear classification with a generalized number of classes can always be decomposed into binary classification of class pairs \cite{hastie2009elements}. Then it is sufficient to consider only binary classification in the sequel.  

\subsubsection{Ideal Case}
Consider the case without channel distortion. The classification margin is illustrated in Fig.~\ref{fig: margin-based classifier} and defined mathematically as follows. Based on the GM model, the classification  margin is defined from the probabilistic perspective. 
To be specific, we firstly define the region boundaries for each cluster, such that the probability that feature vectors, whose ground truths belong to this cluster, cross the region boundaries, is lower than a given threshold denoted as $(1-\zeta)$ with $\zeta \in (0, 1)$. 
Specifically, the probability that the feature vector $\mathbf{x}$ belongs to the ground-truth cluster region defined by $\Theta=\{\mathbf{x}\in\mathbb{R}^D: d_{\mathbf{C}}(\mathbf{x},\bmu)\leq\sigma\}$ where $\sigma>0$, termed \emph{cluster region  radius}, is defined to ensure $\Pr(\mathbf{x}\in\Theta)=\zeta$ and derived in the sequel, where the measure $d_{\mathbf{C}}(\cdot,\cdot)$ denotes the Mahalanobis distance as defined in \eqref{eqn: HSMD}. Hence, the region boundary of the cluster is the isometric surface in terms of the Mahalanobis distance. Then  the corresponding cluster region radius  $\sigma(\zeta)$ characterizes the deviation of cluster points from their centroid given the boundary-crossing probability $(1-\zeta)$. 

\begin{Lemma}[  \cite{gallego2013mahalanobis}]\label{lemma: cluster region}
\emph{Given the GM model, the probability that  an arbitrary  feature vector $\mathbf{x}$ belongs to the ground-truth cluster  region defined by $\Theta=\{\mathbf{x}\in\mathbb{R}^D: d_{\mathbf{C}}(\mathbf{x},\bmu)\leq\sigma\}$ is given as}
\begin{align}
    \Pr(\mathbf{x}\in\Theta)=\frac{\gamma\left(\frac{D}{2},\frac{\sigma^2}{2}\right)}{\Gamma\left(\frac{D}{2}\right)},
\end{align}
\emph{where $\gamma(\cdot,\cdot)$ is the lower incomplete Gamma function defined as $\gamma(s,x)=\int_0^xt^{s-1}e^{-t}dt$, $\Gamma(\cdot)$ is the Gamma function defined as $\Gamma(z)=\int_0^{\infty}t^{z-1}e^{-t}dt$.% and $d_{\mathbf{C}}(\cdot,\cdot)$ denotes the Mahalanobis distance as defined in \eqref{eqn: HSMD}.
}
\end{Lemma}

Combining  Lemma \ref{lemma: cluster region} and $\Pr(\mathbf{x}\in\Theta) = \zeta$, we can obtain the cluster region radius as $\sigma=\sqrt{F_X^{-1}(\zeta;\frac{D}{2},\frac{1}{2})}$, where $F_X^{-1}(\cdot;\frac{D}{2},\frac{1}{2})$ denotes the inverse of the cumulative distribution function of the random variable $X$ following a gamma distribution with shape parameter $\frac{D}{2}$ and rate parameter $\frac{1}{2}$: $X\sim \text{Gamma}(\frac{D}{2},\frac{1}{2})$. 
% As illustrated in Fig. \ref{fig: margin-based classifier}, the region boundaries of two clusters can be  defined as  two parallel  supporting hyperplanes. Following  the literature of statistical learning, the classification margin is  defined as the distance between the two supporting hyperplanes \cite{hastie2009elements}. 
Moreover,  the decision boundary, called separating hyperplanes, lies at    the  middle of supporting hyperplanes as mentioned and denoted as $\mathcal{H}(\mathbf{w},b)$. 
Mathematically, the classification margin, denoted as $\Phi$, is given as 
\begin{align}\label{eqn: margin expression}
    \Phi=\frac{|s(\bmu)|}{\sqrt{\mathbf{w}^T\mathbf{C}\mathbf{w}}}-\sigma=\frac{|s(\bmu)|}{\sqrt{\mathbf{w}^T\mathbf{C}\mathbf{w}}} - \sqrt{ F_X^{-1}\!\!\left(\!\zeta;\frac{D}{2},\frac{1}{2} \right)} ,
\end{align}
where $\frac{|s(\bmu)|}{\sqrt{\mathbf{w}^T\mathbf{C}\mathbf{w}}}$ measures  the Mahalanobis distance of the centroid $\bmu$ from the hyperplane $\mathcal{H}(\mathbf{w},b)$. 

\subsubsection{Case with Channel Distortion}
The result in Lemma~\ref{lemma: cluster region} for the ideal case can be easily extended to the case with channel distortion by replacing  the distance metric $d_{\mathbf{C}}(\cdot,\cdot)$ with the new metric function $d_{\mathbf{C}+\sigma_{\Delta}^2\mathbf{I}}(\cdot,\cdot)$ where $\sigma_{\Delta}^2$ is recalled to be the feature channel error. Specifically, given the same probability constraint $\Pr(\hat{\mathbf{x}}\in\hat{\Theta})=\zeta$, the new cluster region enlarged by channel distortion  can be obtained as  $\Theta'=\{\hat{\mathbf{x}}\in\mathbb{R}^D: d_{\mathbf{C}}(\hat{\mathbf{x}},\bmu)\leq\sigma'\}$, where $\sigma'(\zeta)$ represents the cluster region radius with channel distortion (see Fig. \ref{fig: margin-based classifier}). The channel induced reduction  on the classification is denoted and defined  as $\Delta\Phi=\Phi-\Phi'$, which can be bounded as follow. 
\begin{Proposition}\label{proposition: margin}
\emph{In view of the channel distortion in \eqref{eqn: hat x distribution}, the reduction of the classification margin, i.e., $\Delta\Phi$, can be bounded by}
\begin{align}\label{eqn: margin}
    &\min_d \left\{\frac{\sigma_{\Delta}^2\sigma}{2C_d(C_d+\sigma_{\Delta}^2)}\right\}\mathrm{tr}\left\{\mathbf{C}^{-1}\right\}^{-1}\nonumber\\
    &\quad\leq\Delta\Phi\leq \max_d \left\{\frac{\sigma_{\Delta}^2\sigma}{2C_d(C_d+\sigma_{\Delta}^2)}\right\}\mathrm{tr}\left\{\mathbf{C}+\sigma^2_{\Delta}\mathbf{I}\right\}.
\end{align}
% \emph{where $\tilde{\mathbf{x}}$ is the vector satisfying $d_{\mathbf{C}+\sigma_{\Delta}^2\mathbf{I}}(\tilde{\mathbf{x}},\bmu)=\sigma(\zeta)$ and belongs to the cluster centered at $\bmu$.}
\end{Proposition}
The proof is provided in Appendix~\ref{proof: margin}. The change of the classification margin $\Delta\Phi$ is quantified, and we can observe the relation between margin reduction $\Delta\Phi$ and channel noise variance $\sigma_{\Delta}^2$.
On the one hand, the effect of the feature vector noise lies in reducing the classification margin since the relation $\Delta\Phi\geq0$ can be inferred from the result \eqref{eqn: margin}, resulting in $\Phi'\leq\Phi$.
As the classification margin decreases, the robustness of the classifier deteriorates accordingly. 
On the other hand, the reduction of the margin is an increasing function with respect to the noise variance, i.e., $\sigma_{\Delta}^2$, which is proportional to and comes from the BEP $P_b$.
To prevent the margin from diminishing, it is crucial to control the variance of noise within a specific range.

\subsection{Channel Distortion and Classification Accuracy}\label{Sec: channel distortion and classification accuracy}
The effect of feature channel distortion on classification accuracy is analyzed as follows. 
% Again, we consider binary classification where the two centroids of clusters are $\bmu_0$ and $\bmu_1$.
% We assume that the transmitted feature vector can be perfectly classified, i.e., the two clusters are separable with high probability. (Note: The assumption is relaxed in Section \ref{Sec: Multi-view SemCom}.) 
% A pair of transmitted and received feature vectors should be at the same side of the decision hyperplane of the classifier such that they have the same classified labels. 
% Given the above assumption, this event is referred to as \emph{noisy feature alignment} and its probability gives the correct classification probability. 
% Recall that the received  feature  error vector is distorted by the channel with the error following  the distribution $\Delta\mathbf{x}\sim\mathcal{N}(\mathbf{0},\sigma_{\Delta}^2\mathbf{I})$. 
% Then conditioned on the received feature vector $\hat{\mathbf{x}}$, the correct classification probability is defined as
% $\Pr\left(s(\mathbf{x})s(\hat{\mathbf{x}})>0\mid\hat{\mathbf{x}}\right)$. 
% The correct classification probability is derived as follows. 
To begin with, the distribution of the transmitted feature vector $\mathbf{x}$ conditioned on the received feature vector $\hat{\mathbf{x}}$ is given as 
% \begin{equation}
$    {\mathbf{x}}\mid\hat{\mathbf{x}}\sim\mathcal{N}(\hat{\mathbf{x}},\sigma_{\Delta}^2\mathbf{I}).$
% \end{equation}
Then the distribution of the conditional output score $s({\mathbf{x}})\mid\hat{\mathbf{x}}$ can be derived using the linear relation in \eqref{eqn: output score}. 
The derivation involves the projection of Gaussian distribution onto a particular direction specified by $\mathbf{w}$, which yields the following distribution:
\begin{equation}\label{eqn: score distribution}
    s({\mathbf{x}})\mid\hat{\mathbf{x}}\sim\mathcal{N}(s(\hat{\mathbf{x}}),\sigma_{\Delta}^2).
\end{equation}
Based on the above results, the correct classification probability is presented as follows.
\begin{Lemma}\label{lemma: accuracy}
\emph{Conditioned on the received feature vector $\hat{\mathbf{x}}$, the conditional correct classification probability is given by}
\begin{equation}
    \Pr\!\left(s(\hat{\bf x})s(\mathbf{x})>0|\hat{\bf x}\right)
     = 1-Q\!\left(\!\sqrt{ \frac{3}{(4^n\!-\!1)P_b}}\!\times\!|s(\hat{\mathbf{x}})|\!\right)\!.
\end{equation}
% \emph{where $Q(\cdot)$ is the well-known Q-function defined as $Q(x)=\frac{1}{\sqrt{2\pi}}\int_{x}^{\infty}\exp\left(-\frac{t^2}{2}\right)dt$.}
\end{Lemma}
The proof is provided in Appendix~\ref{proof: accuracy}.
% The proof is straightforward and thus omitted for brevity.
As observed from the above result, a feature vector that is closer to the decision boundary, which is equivalent to a smaller absolute score $|s(\hat{\mathbf{x}})|$, incurs a lower classification accuracy. 
Next, we characterize the classification performance of transmitting a feature vector $\mathbf{x}$ that is randomly sampled from one of the clusters, e.g., class-$\ell$. 
Based on the distribution of the received feature vector presented in \eqref{eqn: hat x distribution}, the corresponding score function has the following distribution:
\begin{align}
    s(\hat{\mathbf{x}})\mid\ell\sim\mathcal{N}\left(s(\bmu_{\ell}), \mathbf{w}^T(\mathbf{C}+\sigma_{\Delta}^2\mathbf{I})\mathbf{w}\right).
\end{align}
Using the result and that in Lemma \ref{lemma: accuracy}, the correct   classification probability  is bounded in the following proposition.
\begin{Proposition}\label{proposition: performance}
\emph{For a feature vector $\mathbf{x}$ sampled from the GM model of two clusters, the correct classification accuracy of the received vector $\hat{\mathbf{x}}$ is lower bounded by}
% \begin{align}
%     \Pr\left(s(\hat{\bf x})s(\mathbf{x})>0\right)\geq 1-\frac{1}{4}\sqrt{\frac{(4^n-1)P_b}{2(4^n-1)P_b+3\mathbf{w}^T\mathbf{C}\mathbf{w}}}\sum_{\ell=0}^1\exp\left(-\frac{s(\bmu_{\ell})^2}{2\mathbf{w}^T\mathbf{C}\mathbf{w}}\right).
% \end{align}
\begin{align}
    &\Pr\left(s(\hat{\bf x})s(\mathbf{x})>0\right)\geq\nonumber\\
    &1\!-\!\sqrt{\!\frac{(4^n\!\!-\!\!1)P_b}{2(4^n\!\!-\!\!1)P_b\!+\!3\mathbf{w}^T\!\mathbf{C}\mathbf{w}}}\exp\!\!\left(\!-\frac{3(\sigma\!+\!\Phi)^2\mathbf{w}^T\!\mathbf{C}\mathbf{w}}{2\left(2(4^n\!\!-\!\!1)P_b\!+\!3\mathbf{w}^T\!\mathbf{C}\mathbf{w}\right)}\!\right)\!.
\end{align}
\end{Proposition}
The proof is provided in Appendix \ref{proof: performance}. 
We make the following two  observations from the above result. 
\begin{itemize}
    \item The term in the exponent, i.e., $(\sigma+\Phi)^2=\frac{s(\bmu_{\ell})^2}{\mathbf{w}^T\mathbf{C}\mathbf{w}}$, represents the squared Mahalanobis distance from the centroid $\bmu_{\ell}$ to the decision hyperplane $\mathcal{H}(\mathbf{w},b)$.
    The larger the distance is,  the more separable  the two clusters and thus the wider the margin, i.e., $\Phi$, and vice versa. 
    Correspondingly, the correct classification probability $\Pr\left(s(\hat{\bf x})s(\mathbf{x})>0\right)\to1$ diminishes as the said distance grows.
    \item The derived result is consistent with the intuition that the  correct classification probability should be a monotone decreasing function of the BEP.
    In particular, from its lower bound,  $\Pr\left(s(\hat{\bf x})s(\mathbf{x})>0\right)$ is found to scale with  $P_b$ as  $\Pr\left(s(\hat{\bf x})s(\mathbf{x})>0\right)=1-\mathcal{O}(\sqrt{P_b})$.
\end{itemize}

% \subsection{Classification Margin Enhancement}\label{Subsection: Margin Enhancement}
\section{Classification Margin Enhancement}\label{Subsection: Margin Enhancement}
Due to either channel or sensing noise, single-transmission or single-view robotic observations may not achieve the level of classification accuracy required for accurate KP identification. In this section, we discuss and quantify  methods to enhance the classification margin and thus the classification accuracy for object detection. 
\subsection{Margin Enhancement Approaches}
Different methods for classification margin enhancement largely belong to one of the following two approaches. 
\begin{itemize}
    \item \emph{Channel distortion reduction:} Reducing channel distortion enlarges the effective classification margin as shown in Proposition \ref{proposition: margin}. Two popular methods for the purpose are  \emph{code rate reduction} and \emph{retransmission} both at the cost of increased latency. 
    The code rate refers to the ratio of information bits and the total number of transmitted bits. Reducing the rate provides more redundancy for  error correction and hence reduces the variance of feature channel distortion  (i.e., $\sigma_{\Delta}^2$).  
    On the other hand, retransmission, a reliable-communication technique widely adopted in practice,  can increase the link reliability by exploiting channel temporal diversity.  of the communication link by providing additional opportunities to successfully transmit data that may have been corrupted during the initial transmission. 
    Consequently, both techniques reduce the variance of the channel perturbed error vector, and thus enlarge the classification margin.
    \item \emph{Cluster variance reduction:} An alternative method for enhancing the effective margin is  \emph{multi-view feature aggregation} \cite{liu2023over}. 
    By averaging or maximizing the feature vectors obtained from  different views, the variances of data clusters are reduced to enhance the classification accuracy by enlarging the classification margin. In this section, we consider the average pooling of feature vectors extracted from robot's observations of an object from different angles or locations.
\end{itemize}

\subsection{Multi-view Robotic SemCom}\label{Sec: Multi-view SemCom}
The accuracy enhance via multi-view classification is mathematically characterized as follows.   Let $m$ denote the number of observations for each object. If an object is observed for $m$ times, the aggregated  feature vector is $\overline{\hat{\mathbf{x}}}=\frac{1}{m}\sum_{i=1}^m\hat{\mathbf{x}}(i)$, which suppresses the cluster variance by $m$ times. Thus, using  \eqref{eqn: PDF of noisy feature vector}, the PDF of $\overline{\hat{\mathbf{x}}}$ is obtained as 
\begin{equation}
    f(\overline{\hat{\mathbf{x}}}|m)=\frac{1}{L}\sum_{\ell=1}^L\mathcal{N}\left(\overline{\hat{\mathbf{x}}}~\Big|~\bmu_{\ell},\frac{1}{m}(\mathbf{C}+\sigma_{\Delta}^2\mathbf{I})\right).
\end{equation}
Following the same procedure for single-view analysis, the correct classification probability for multi-view classification is characterized as follows.
\begin{Proposition}\label{proposition: multi-view}
\emph{Assume that  multi-view feature vectors are independently generated using the binary GM model with two clusters. Let $\overline{\mathbf{x}}$ denote the aggregated feature vector  of  $m$ views of an arbitrary  object,  the correct classification accuracy is lower bounded as follows:}
\begin{align}
    &\Pr\left(s(\overline{\hat{\bf x}})s(\overline{\mathbf{x}})>0|m\right)\geq\nonumber\\
    &1\!-\!\sqrt{\!\frac{(4^n\!-\!1)P_b}{2(4^n\!\!-\!\!1)P_b\!+\!3\mathbf{w}^T\!\mathbf{C}\mathbf{w}}}\exp\!\!\left(\!-\frac{3m(\sigma\!+\!\Phi)^2\mathbf{w}^T\!\mathbf{C}\mathbf{w}}{2\left(2(4^n\!\!-\!\!1)P_b\!+\!3\mathbf{w}^T\!\mathbf{C}\mathbf{w}\right)}\!\right)\!.
\end{align}
\end{Proposition}
The proof is provided in Appendix \ref{proof: multi-view}.
The probability is observed to approach the maximum of $1$ at an exponential rate with respect to the number of views: $\Pr\left(s(\overline{\hat{\bf x}})s(\overline{\mathbf{x}})>0|m\right)=1-\mathcal{O}(e^{-m})$. This is much faster than the sub-linear scaling law with respect to  the BEP that $\Pr\left(s({\hat{\bf x}})s(\mathbf{x})>0|m\right) = 1-\mathcal{O}(\sqrt{P_b})$. Therefore, we can conclude that multi-view observations is a more effective approach than retransmission discussed the following sub-section  to enhance the classification accuracy.

\subsection{Feature Vector Retransmission}\label{Sec: Retransmission}
Next, we consider retransmission and derive the required number of retransmissions of a feature vector to achieve the target classification  accuracy $\xi$. 
In this scheme, the received  feature vector results from averaging over one initial transmission and $(T-1)$ retransmissions, i.e., 
$\hat{\mathbf{x}}(T)=\frac{1}{T}\sum_{t=1}^T\hat{\mathbf{x}}(t)$,
where $\hat{\mathbf{x}}(t)$ refers to the received feature vector during the $t$-th transmission. The averaging suppresses the error variance as follows: 
$\Delta\mathbf{x}\sim\mathcal{N}\left(\mathbf{0},\frac{\sigma_{\Delta}^2}{T}\mathbf{I}\right)$.
Given $\xi$, the required   transmission  latency is characterized in the following proposition.
\begin{Proposition}\label{proposition: retransmission}
\emph{To achieve  the target classification accuracy, $\xi$, the required  number of transmissions for each  feature vector  is given  as}
% \begin{equation}\label{eqn: retransmission}
%     T=\left\lfloor\frac{(4^n-1)P_b}{3\mathbf{w}^T\mathbf{C}\mathbf{w}}\left\{\frac{1}{16(1-\xi)^2}\left[\sum_{\ell=0}^1\exp\left(-\frac{s(\bmu_{\ell})^2}{2\mathbf{w}^T\mathbf{C}\mathbf{w}}\right)\right]^2-2\right\}\right\rfloor^+,
% \end{equation}
\begin{equation}\label{eqn: retransmission}
    T\!=\!\left\lceil\!\frac{(4^n\!\!-\!1)P_b}{3\mathbf{w}^T \mathbf{C}\mathbf{w}}\!\left(\!\frac{2(\sigma + \Phi)^2}{\mathcal{W}\!\left( 2(\sigma\!+\!\Phi)^2e^{(\sigma\!+\!\Phi)^2} (1\!-\!\xi)^2 \right)}\!-\!1\!\right)\!\right\rceil\!\!,
\end{equation}
\emph{where $\lceil\cdot\rceil$ is the ceiling function, defined as $\lceil x\rceil=\min\{n\in\mathbb{Z}\mid n\geq x\}$, and $\mathcal{W}(\cdot)$ is the Lambert $W$ function.}
\end{Proposition}
The proof is provided in Appendix \ref{proof: retransmission}.
% The result  straightforwardly follows from Proposition \ref{proposition: performance} and its proof is omitted for brevity.
Two  observations can be made from the above result. 
First, consider the cases where  the data set exhibit high discernability as determined by a sufficiently large  classification margin $\Phi$, and/or if the BEP $P_b$ is low. In such cases, the value within the floor function in \eqref{eqn: retransmission} is less than $1$ to make retransmission unnecessary. 
Next, if the communication link is highly unreliable as reflected in a high BEP $P_b$, or  the separation between the two centroids is insufficient  to mitigate channel distortion,  retransmissions are needed and their required number  is observed to be  a monotone increasing function of the target classification accuracy  $\xi$.

\section{Experimental Results}
% \hl{xxx: After finish, please revise 2nd bullet of contributions accordingly. }

\subsection{Experimental Settings}
% The experimental setups are as follows unless specified otherwise. 

\begin{itemize}

\item {\bf Linear classifier with synthetic data:} We consider linear classifiers trained on data features following the GM distribution. The dimensionality of the feature space is set as $D=100$.  Each feature is represented by a signed integer with $n=8$ bits (i.e., Int8). For both the cases of a small (i.e., binary) and a large (i.e., 10) numbers of object classes, we assume features are i.i.d. and the covariance matrix is diagonalized through PCA.
In the binary GM model, one cluster centroid is the vector with all elements being $+1$, while the other is the vector with all  elements being $-1$. 
In the 10-class GM model, these 10 data clusters share the same identity covariance matrix. The centroids of the 10 clusters are specified as follows: for class-$\ell$, the elements from dimension $10(\ell-1)$ to dimension $(10\ell-1)$ are set to $-1$, while the remaining dimensions are set to $+1$.
The linear classifier operates by excluding one label at a time by binary classification, and outputs the final remaining label.

\item {\bf DNN classifier with real data:} We construct different KPs using  the multi-view ModelNet40 dataset that comprises $40$ categories of common three-dimensional household objects \cite{su2015multi}.
% There are in total $40$ objects existing in the robot's environment, and multiple KPs are randomly generated with each containing $5$ objects. 
% \hl{xxx: What are the relations between the KPs in terms of same/differnt objects?} 
% For every pair of constructed KPs, there is one overlapped object.
% \hl{xx: How many are task relevant objects?}
% Hence, there are 20 task relevant objects in total.
For each object, the robot can sense up to $12$ views.
The DNN model used at the server is a VGG-11 model trained on multi-view ModelNet40 where the fusion module is an average pooling layer.
The output of the final classifier is passed through a softmax layer to obtain the prediction confidence score for DNN model, or say, the prediction logits, and thus the scores summed over $40$ categories is 1.
The inferred label is associated with the class with the highest score.
% To guarantee the prediction accuracy, the default logit threshold is set to be $0.6$, which far exceeds $1/40$ corresponding to uniformly random guess.

\item {\bf Communication  model:} 
We consider a Rician fading channel and apply channel inversion technique to ensure fast transmission \cite{tse2005fundamentals}. The bandwidth is set as $B=1$ MHz. We consider two distinct uncoded transmission schemes as our proposed ULL-FT in contrast to the conventional reliable transmission using error control codes. The two schemes are described as follows:
\begin{itemize}
    \item[1) ] \underline{Scheme 1:} The modulation is fixed as \emph{binary phase shift keying} (BPSK), resulting in a constant transmission rate of 1 Mbits/s. In this context, given the BEP, the channel SNR can be directly established, enabling an effective comparison with its reliable transmission counterpart. We assess this scheme as ULL-FT by conducting simulations with given BEPs.
    \item[2) ] \underline{Scheme 2:} We apply adaptive \emph{quadrature amplitude modulation} (QAM), which leads to variable BEPs and transmission rates subject to different SNRs. This scheme tends to be identical to Scheme 1 when the SNR is low. We evaluate this scheme as ULL-FT by performing simulations with given SNRs. 
\end{itemize}
\item {\bf Task-relevant KPs model:} We design a task as follows: task-relevant objects are randomly selected from the total object set to generate multiple KPs, each of them consists of different task-relevant objects. 
As for the total objects in the environment, 80 percent of them are assigned as task-relevant while the remaining of them are irrelevant.
Then, we establish several distinct KPs comprising multiple task-relevant objects under different settings, e.g., the task-relevant KPs, represented as $\text{PathLength}=[10~10~10]$, consist of three KPs, each containing 10 objects. 

\item {\bf Object-exploration model:}
% We design a certain task as follows: task-relevant objects are randomly selected from the total object set to generate multiple KPs, each of them is composed of different clusters of task-relevant objects. 
% As for the total objects, 80 percent of them are assigned as task-relevant while the remaining 20 percent of them are irrelevant.
% Then, we establish several distinct KPs comprising multiple task-relevant objects under different settings. 
% When an object is detected by the robot, the extracted features undergo a binary classification: either it is classified as an irrelevant object, with a predefined probability denoted as $p_I = 0.2$, or it can be associated with one of the KPs with a prior probability of 0.8.
The robotic exploration of objects is modeled as a Poisson arrival process characterized by an arrival rate parameter $\lambda = v/\bar{d}$, which characterizes the number of objects explored by the robot within a unit of time. Here, $v$ corresponds to the robot's mobility and $\bar{d}$ is the average separation distance between objects. 
For objects that are identified as task-relevant, the subsequent step involves determining which specific KP they belong to. The task is deemed completed as long as one of the task-relevant KPs has been identified, i.e., the objects in this KP has been found via classification to exist in the environment.
% \item  {\bf Path-hitting Latency:}
% Consider arbitrary object set $\mathcal{T}_t$ of the $t$-th KP. 
% The task completion along KP requires that all affiliated objects should be hit by robots. To measure the performance of hit events along KP, we first introduce the probability of all objects of the $t$-th KP being hit at least once, called \emph{path-hitting probability}, denoted as $p_{t}$. 
% Accordingly, the \emph{path-hitting latency} is defined as the latency required for all objects of the $t$-th KP being observed by the robot at least once.
\item {\bf Reliable transmission benchmark}: 
% The channel conditions are characterized by the BEP, which results from uncoded BPSK transmission. Specifically, when the BEP is $P_b$, the energy per bit to noise power spectral density ratio ($E_b/N_0$) can be calculated from $E_b/N_0=(Q^{-1}(P_b))^2/2$. 
% The spectrum efficiency of BPSK is 0.5 bits/s/Hz, so that the bit rate is $r_b=0.5B$ bits/s, where $B$ represents the bandwidth.
% For uncoded transmission, the information bit rate is equal to the gross bit rate. Consequently, the carrier-to-noise ratio, i.e., SNR of the received signal, can be determined as a function of the BEP: $\gamma=\frac{E_b}{N_0}\frac{r_b}{B}=Q^{-1}(P_b))^2/4$. 
Aligned with 6G URLLC requirements, the scheme adopts channel coding with finite block length to achieve an ultra-high reliability with BEP of $10^{-9}$. 
Let $R$ denote the \emph{coding rate} defined as the ratio of the number of feature vector bits, i.e., $nD$ as mentioned earlier, and the number of transmitted symbols, denoted as $N$. 
Then the SNR $\gamma$, packet error probability $\varepsilon$, coding  rate $R$, and block length $N$ can be related using the following well-known result from information theory (see e.g., \cite{polyanskiy2010channel}) 
\begin{equation}\label{eqn: short packet}
    R(\gamma,\varepsilon)\!=\!\frac{nD}{N}\!=\!C(\gamma)-\sqrt{\frac{V(\gamma)}{N}}Q^{-1}(\varepsilon)+\frac{\log_2N}{2N},
\end{equation}
where $C(\gamma)$ represents the channel capacity and $V(\gamma)$ denotes the channel dispersion \cite{polyanskiy2009dispersion}. 
Then, the minimum $N$ can be determined from solving \eqref{eqn: short packet} numerically.
This gives the communication latency $T=N/B$ since the symbol duration is $1/B$.
\end{itemize}
% In this benchmark scheme, reliable transmission refers to the block decoding error probability, denoted by $\varepsilon$, below the threshold of $10^{-9}$. To achieve this threshold, redundancy is incorporated into the transmission of feature vector through error control codes. Given channel condition, characterized by SNR $\gamma$, and the reliabilility requirement, represented by block error probability $\varepsilon$, the coding rate can be determined from \cite{polyanskiy2010channel} as follows:
% \begin{equation}\label{eqn: short packet}
%     R(\gamma,\varepsilon)=C(\gamma)-\sqrt{\frac{V(\gamma)}{N}}Q^{-1}(\varepsilon)+\frac{\log_2N}{2N},
% \end{equation}
% where $C(\gamma)=\frac{1}{2}\log_2(1+\gamma)$ represents the channel capacity and $V(\gamma)$ denotes the channel dispersion \cite{polyanskiy2009dispersion}, which is given by $V(\gamma)=\frac{\gamma}{2}\frac{\gamma+2}{(\gamma+1)^2}(\log_2e)^2$. 
% Considering the feature vector consists of $nD$ bits and the available bandwidth is $B$ Hertz,  the latency can be calculated by $T=\frac{nD}{R}\cdot\frac{1}{B}$ seconds.

\subsection{Robotic SemCom with Linear Classification}
\subsubsection{Performance of ULL-FT}
% \begin{figure}[t!]
% \centering
% \subfigure[Accuracy vs BEP]{
% \includegraphics[width=0.48\textwidth]{Accuracy_vs_BER.pdf}}
% \subfigure[Latency vs BEP]{
% \includegraphics[width=0.48\textwidth]{Latency_vs_BER.pdf}}
% \caption{The effects of BEP on the performance. (a) The effects of the BEP on the classification accuracy in terms of different discriminant gains. (b) The comparison between the reliable transmission and ultra-low-latency transmission about the transmission latency of one feature vector under different channel conditions.}
% \label{Accuracy vs BER}
% \end{figure}
\begin{figure}[t!]
\centering
\subfigure[Binary classification accuracy]{
\includegraphics[width=0.7\columnwidth]{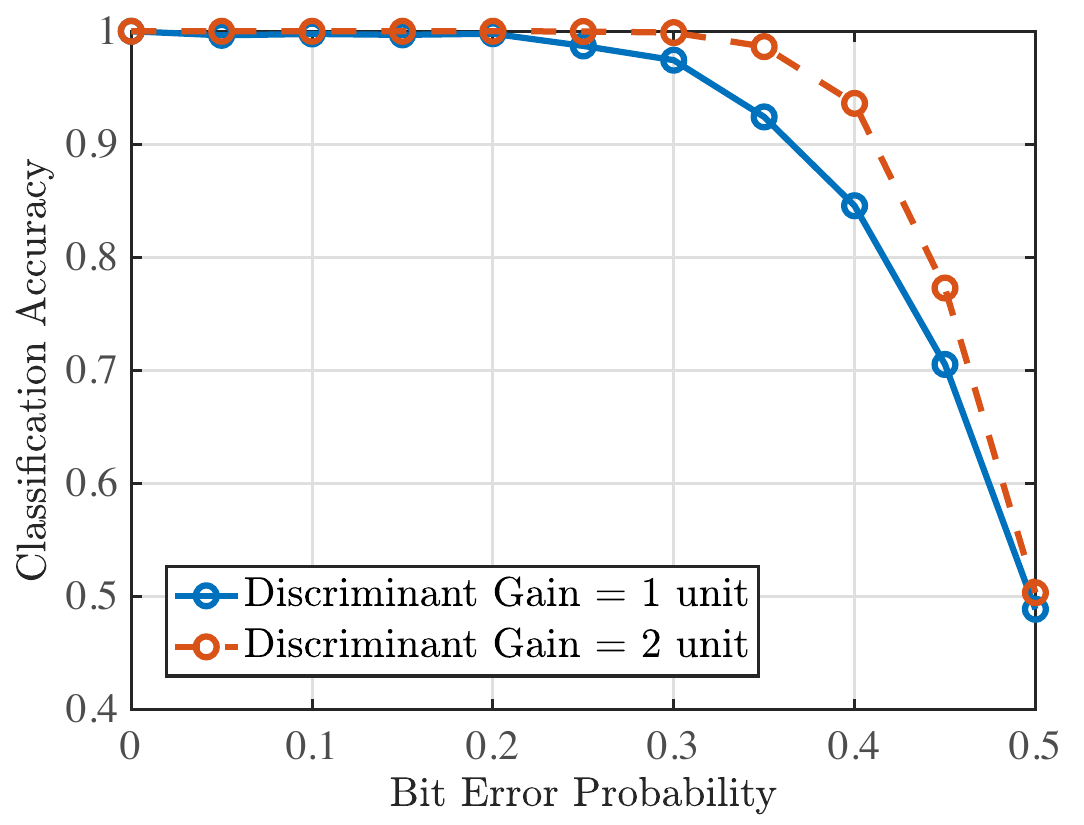}\label{Accuracy vs BER (binary)}}
\subfigure[10-class classification accuracy]{
\includegraphics[width=0.7\columnwidth]{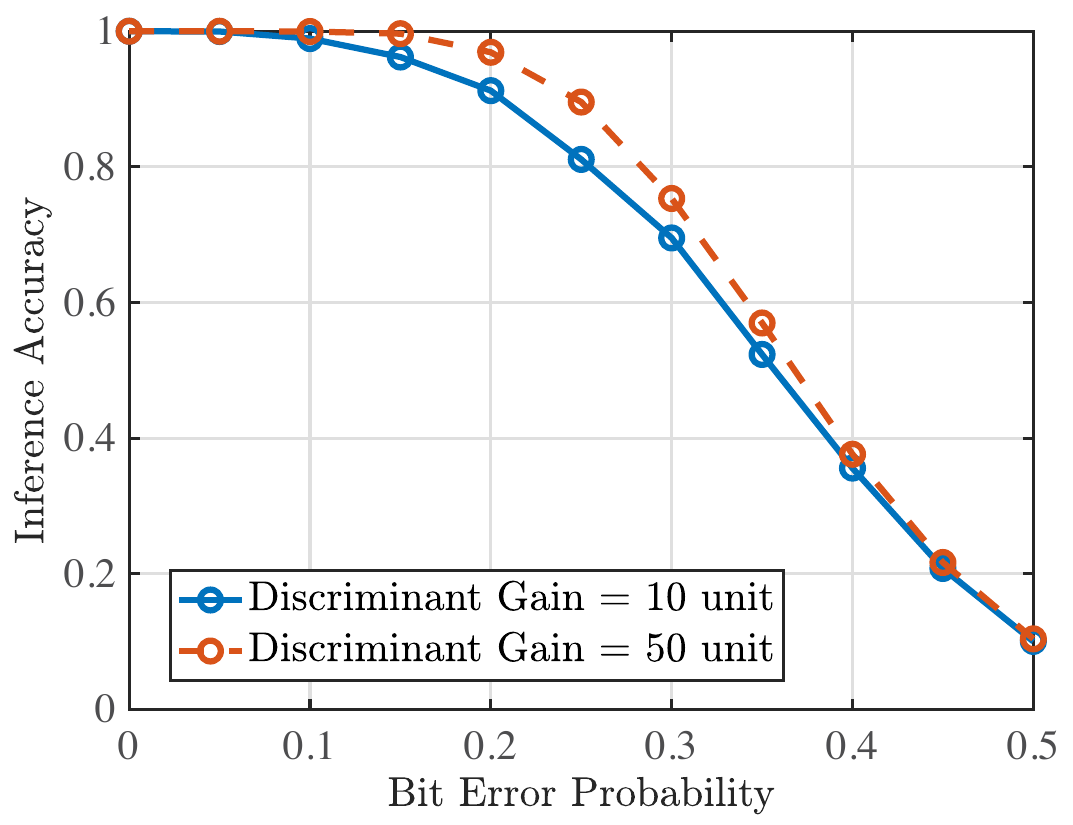}\label{Accuracy vs BER (10-class)}}
% \subfigure[Transmission latency]{
% \includegraphics[width=0.7\columnwidth]{Simulations/GMM_Latency_BEP.pdf}\label{Latency vs BER}}
\caption{Accuracy and latency of ULL-FT scheme. The effects of BEP on the classification accuracy for different discriminant gains and for (a) binary classification or (b) 10-class classification.}
\label{Exp: Ultra-Low-Latency}
\end{figure}

\begin{figure}[t!]
\centering
\subfigure[Transmission latency with different BEPs]{
\includegraphics[width=0.72\columnwidth]{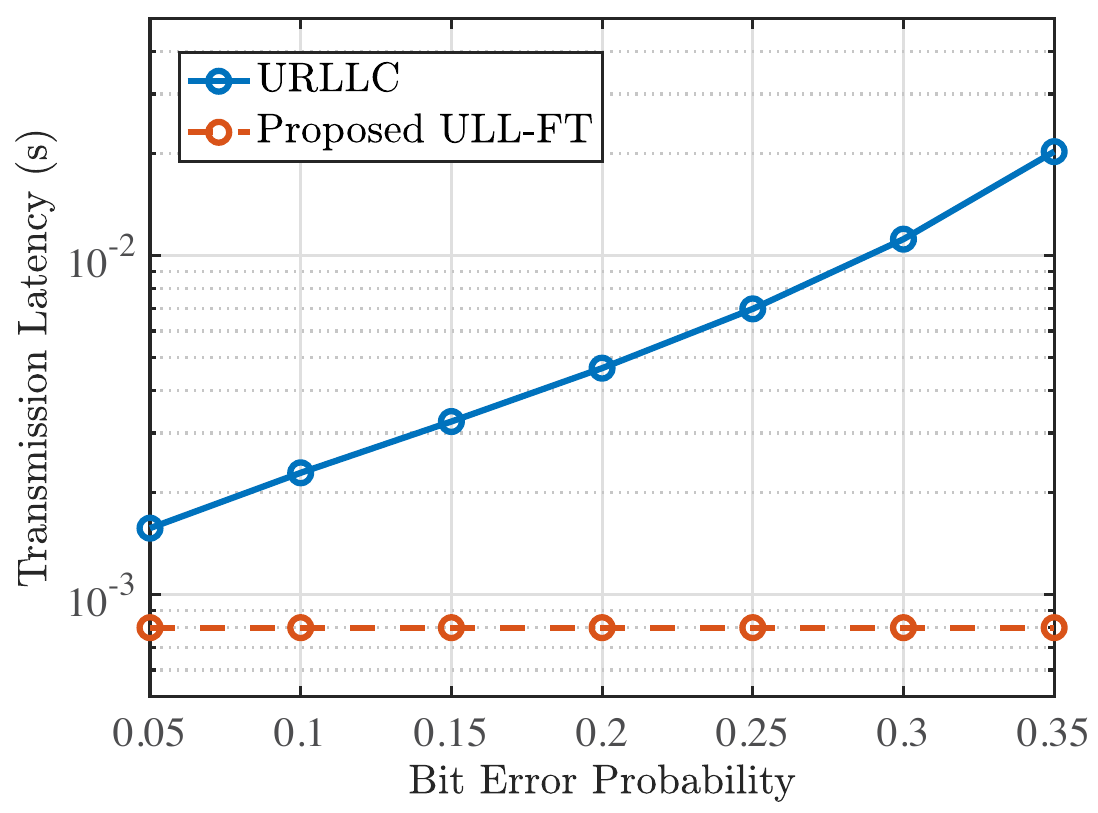}\label{Latency vs BER}}
\subfigure[Transmission latency under different SNRs]{
\includegraphics[width=0.72\columnwidth]{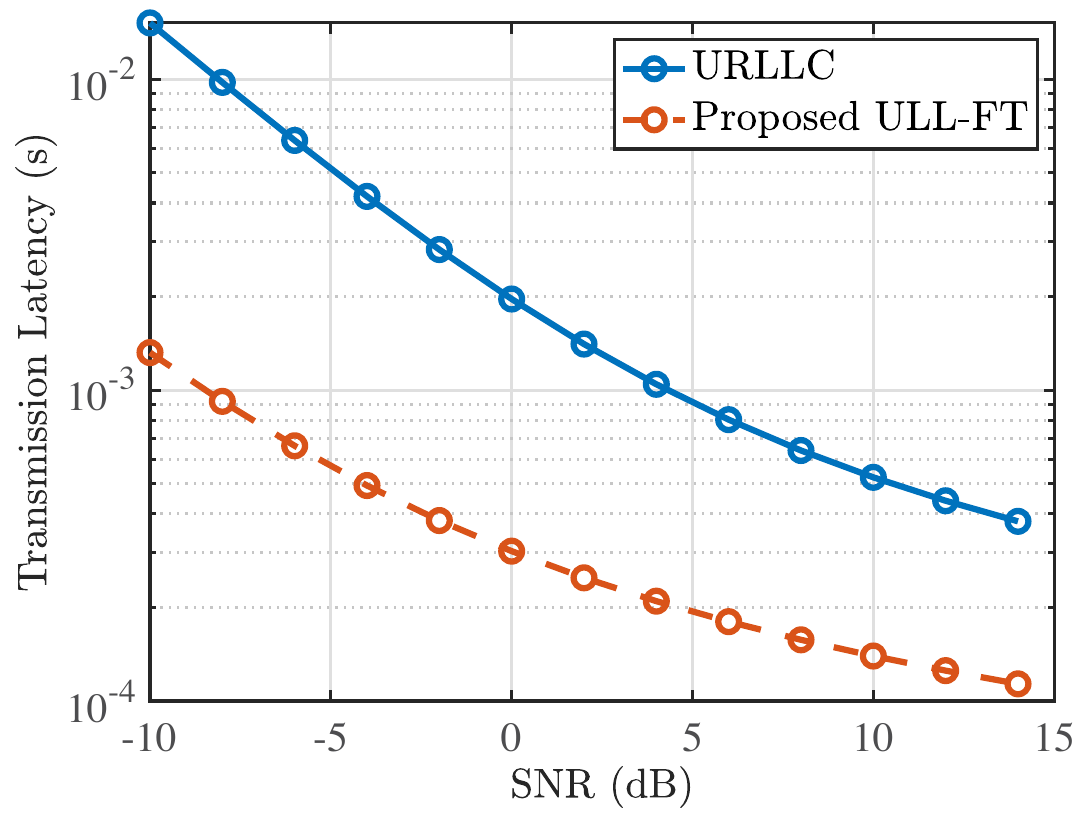}\label{Latency vs SNR}}
\caption{Transmission latency comparison between ULL-FT and classic URLLC. The proposed ULL-FT refers to (a) BPSK in Scheme-1 and (b) adaptive QAM in Scheme-2, respectively.}
\label{Exp: Ultra-Low-Latency}
\end{figure}

The discriminant gain, representing the distance between centroids of the two clusters as defined in \eqref{eqn: discriminant gain}, is measured in the unit of 20. 
The curves of classification accuracy with ULL-FT versus BEP are plotted in Fig. \ref{Accuracy vs BER (binary)} and Fig. \ref{Accuracy vs BER (10-class)} corresponding to binary and 10-class classification, respectively. Different discriminant gains are considered. 
% While multi-class classification poses a greater challenge compared to the binary case, both exhibit similar robustness behavior, as depicted in Fig. \ref{Accuracy vs BER (binary)} and Fig. \ref{Accuracy vs BER (10-class)}, respectively.
% Therefore, we will use Fig. \ref{Accuracy vs BER (binary)} as an illustrative example to discuss the following noteworthy observations:
For all cases, one can observe the existence of a BEP range within which classification accuracy remains constant. 
However, increasing the BEP beyond such a range triggers rapid deterioration of classification accuracy.
This observation indicates the existence of classification margin and hence the classification model's robustness to withstand a certain level of channel distortion. 
The broader the BEP range with constant accuracy, the larger the classification margin.
In this aspect, increasing the discriminant gain is observed to enlarge the margin.
Note that the accuracy at low BEP corresponds to the performance of traditional reliable transmission.

Next, we examine ULL-FT under the designation of Scheme-1. For both ULL-FT and reliable transmission for transmitting a single feature vector, the curves of transmission latency versus BEP are plotted in Fig. \ref{Latency vs BER}.
Due to uncoded transmission, the latency of ULL-FT is independent of BEP.
On the contrary, in the case of reliable transmission, coding redundancy is needed to achieve transmission reliability and can be excessive at a high BEP.
As a result, the latency grows rapidly as BEP increases as observed from Fig. \ref{Latency vs BER}. This stymies ultra-low-latency communication desired for robotic SemCom.
As an example for comparison, let us consider the case in Fig. \ref{Accuracy vs BER (binary)} with discriminant gain of 2 units. A BEP below 0.35 achieves the classification accuracy up to 0.97. 
Then, with similar classification performance at BEP $=0.35$, the transmission latency of reliable transmission is close to two-order of magnitude higher than that of ULL-FT.

Lastly, we evaluate ULL-FT under the framework of Scheme-2. Performance assessments of ULL-FT are conducted through simulations under varying effective channel SNRs, and it is then directly compared to the URLLC benchmark. As illustrated in Fig.~\ref{Latency vs SNR}, a pattern becomes evident: the feature transmission latency for both reliable transmission and the ULL-FT scheme decreases as the channel SNR increases. Compared to URLLC scheme, ULL-FT exhibits a remarkable decrease in latency, particularly at lower SNRs. For instance, when SNR equals 10dB, the resulting BEP for ULL-FT is 0.15, which falls within the margin of accommodation, ensuring no performance loss in classification accuracy. Meanwhile, it records a latency level an order of magnitude lower than that of its URLLC counterpart. This significant reduction in latency underscores the effectiveness of the ULL-FT scheme.

\subsubsection{Robotic SemCom with Retransmission and Multi-view}

\begin{figure}[t!]
\centering
\subfigure[Classification accuracy with retransmission]{
\includegraphics[width=0.7\columnwidth]{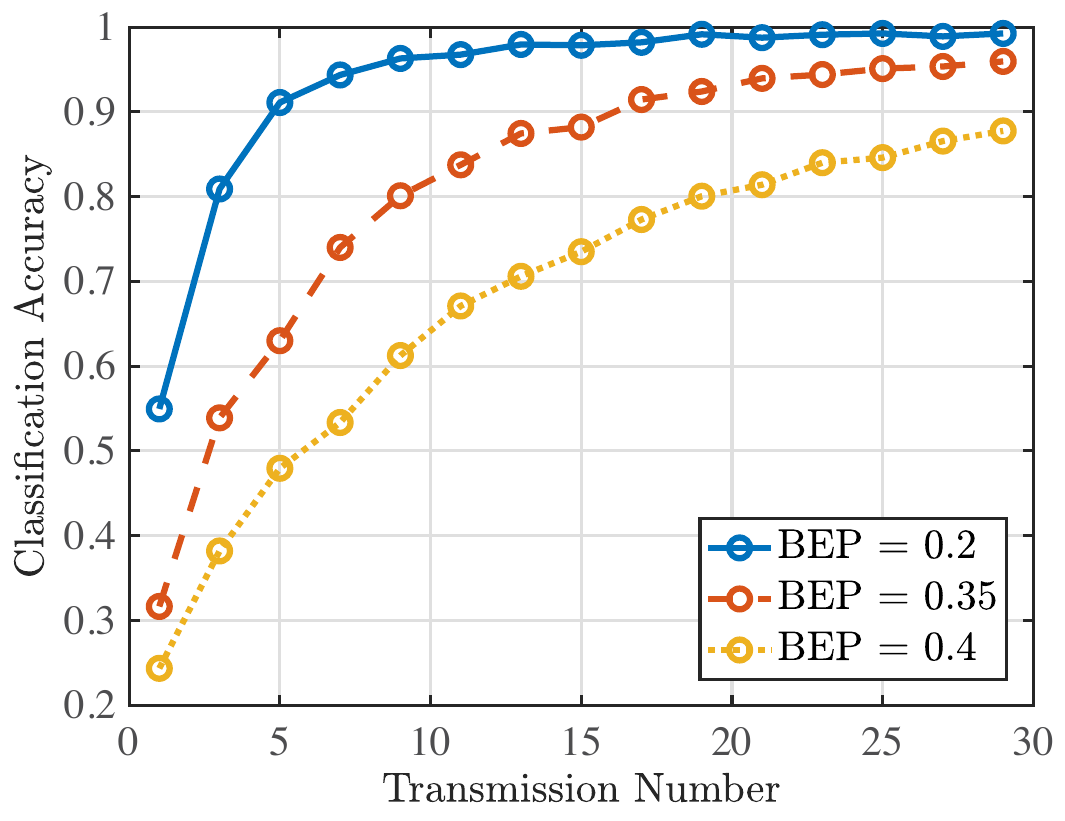}\label{Fig: Retransmission}}
\subfigure[Classification accuracy with multiple views]{
\includegraphics[width=0.7\columnwidth]{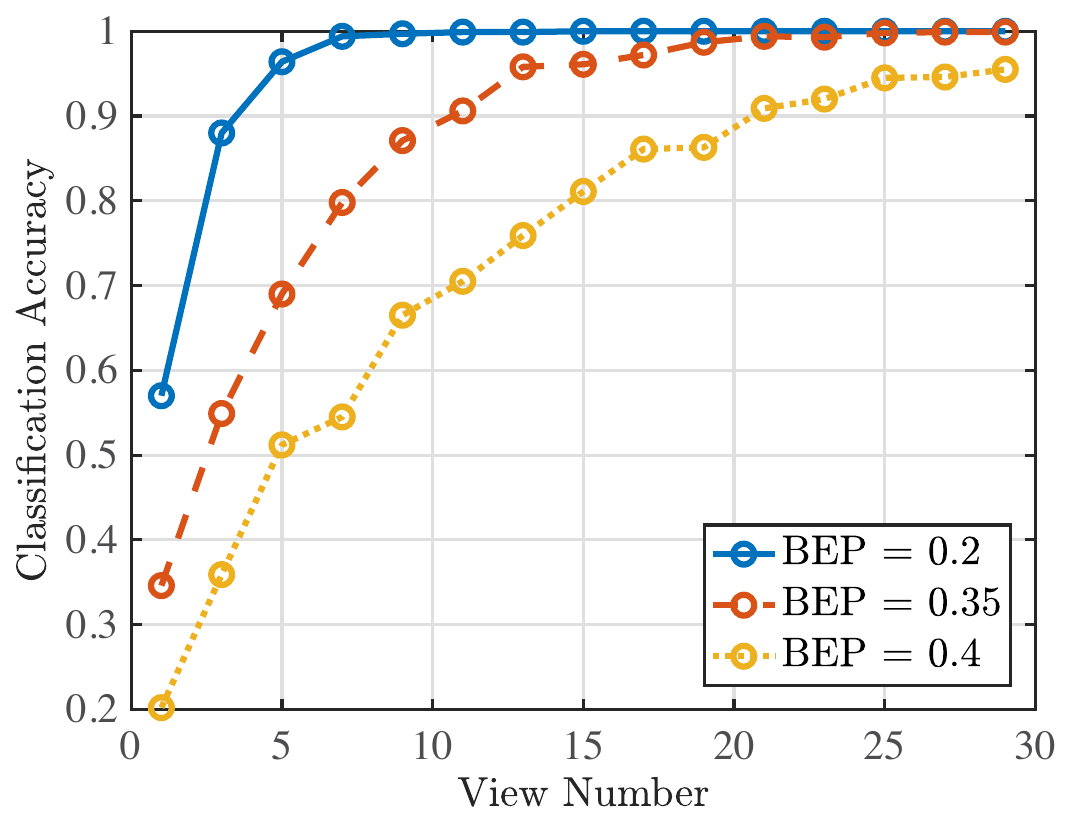}\label{Fig: Accuracy vs Number of views}}
\caption{The performance of robotic SemCom with (a) retransmission and (b) multi-view classification.}
\label{Fig: multi-view}
\end{figure}

Consider the 10-class classification with classification margin enhanced by either retransmission or case of multi-view.
The curves of classification accuracy versus retransmission number or view number are depicted in Fig. \ref{Fig: Retransmission} and Fig. \ref{Fig: Accuracy vs Number of views}, respectively. 
Different BEPs are considered.
For all cases, the accuracy is observed to consistently improves at the cost of growing communication overhead (hence latency) via either retransmission or multi-view uploading.
One can observe that at a relatively low BEP (e.g., 0.2), the improvement rate is higher, allowing the accuracy to reach its maximum with relatively small increase in communication overhead. 
On the contrary, at extremely high BEP (e.g., 0.4), the rate is much lower.
Overall, the results confirm the effectiveness of classification margin enhancement techniques.

\begin{figure}[t!]
\centering
\subfigure[Robotic transmission latency with different BEPs]{
\includegraphics[width=0.71\columnwidth]{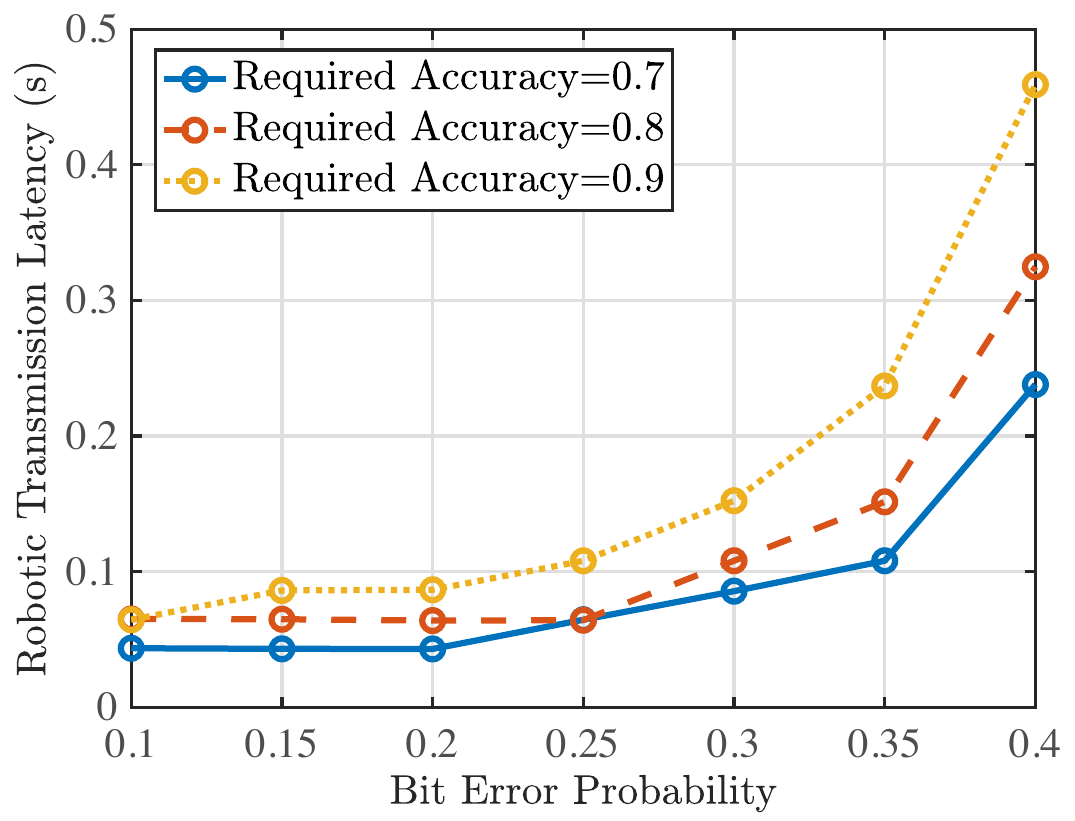}\label{Fig: Number of views vs BER}}
\subfigure[Robotic transmission latency under different SNRs]{
\includegraphics[width=0.71\columnwidth]{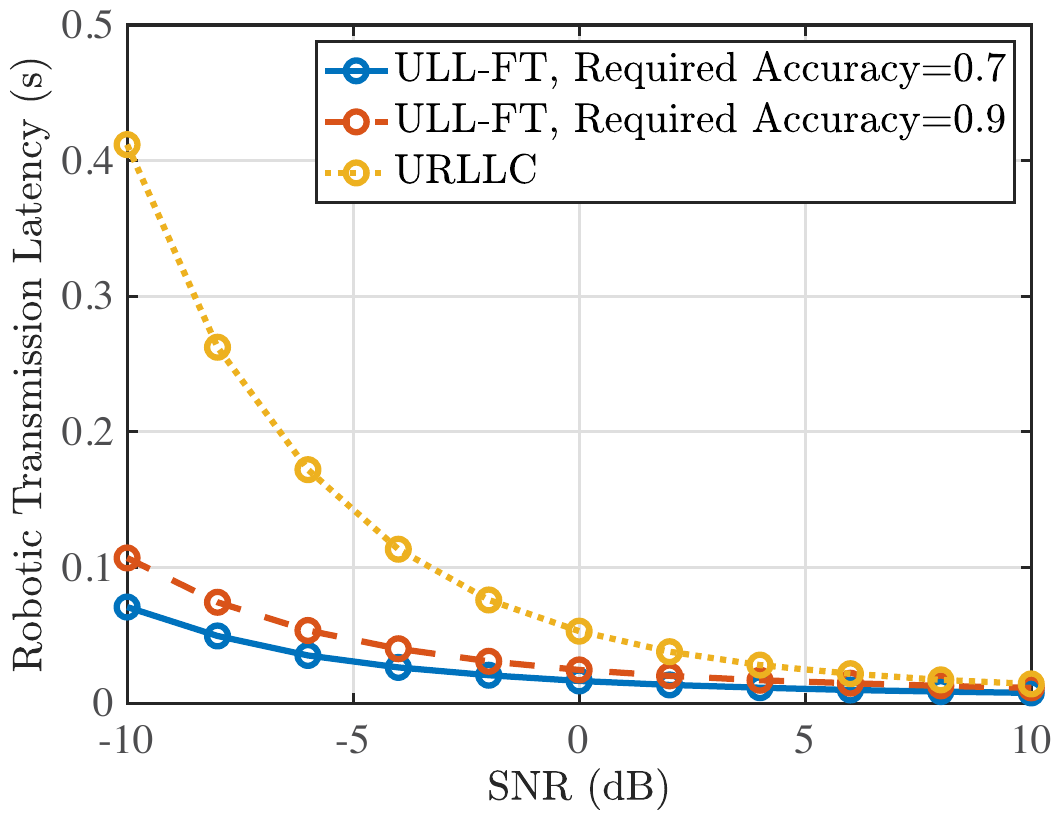}\label{Fig: Number of views vs SNR}}
\caption{The robotic SemCom transmission latency in the robotic exploration process for different classification accuracy requirements. ULL-FT is implemented by (a) Scheme-1 with different BEPs and (b) Scheme-2 under different SNRs.}
\end{figure}
\subsubsection{Robotic Transmission Latency}
Consider the 10-class classification with discriminant gain of 1 unit. 
The robotic exploration process ends upon the occurrence of a path-hitting event, where the server identifies all relevant objects in one of the task-relevant KPs. 
The task-relevant KPs, represented as $\text{PathLength}=[10~10~10]$, consist of three KPs, each containing 10 objects. Each time the robot encounters an object, it observes the object multiple times until the desired accuracy is achieved. This process continues until one of the task-relevant KPs has been identified (i.e., hitting a KP).
Then the robotic transmission latency is determined by both the encountered objects and the number of observations for each object during the process.
Scheme-1 is depicted in Fig. \ref{Fig: Number of views vs BER}, where the curves of robotic transmission latency during exploration versus BEP are depicted for different accuracy requirements for object recognition.
Reflecting the advantage of ULL-FT via classification margin exploitation, the robotic transmission latency during exploration grows only gradually as BEP increases except for when it is extremely high (i.e., larger than 0.3).
The above observation is similar for different accuracy requirements.
Scheme-2 is applied in Fig.~\ref{Fig: Number of views vs SNR}, where the ULL-FT scheme has significantly lower robotic transmission latency during exploration than the URLLC under different SNRs. The difference is particularly noticeable at lower SNRs. Furthermore, lowering the required accuracy can further decrease the latency.

\begin{figure}[t!]
\centering
\subfigure[Top-1 accuracy with single-view classification]{
\includegraphics[width=0.7\columnwidth]{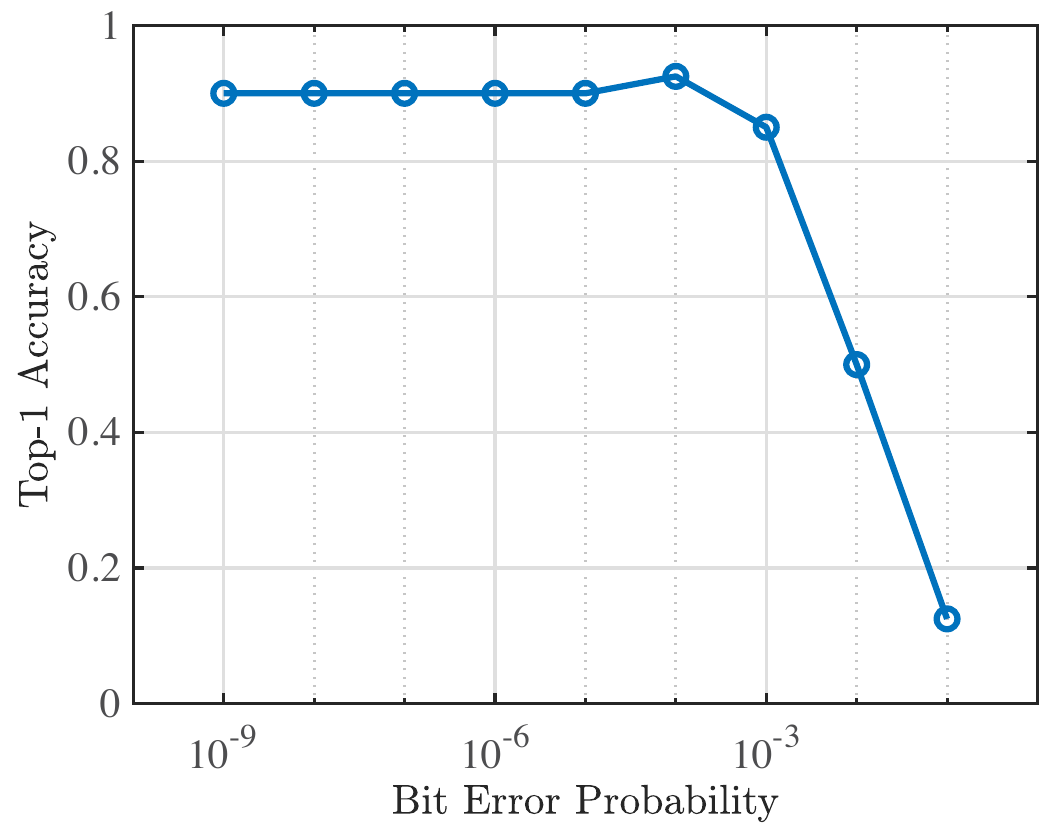}\label{Fig: DNN Accuracy vs BER}}
\subfigure[Top-1 accuracy with multi-view classification]{
\includegraphics[width=0.7\columnwidth]{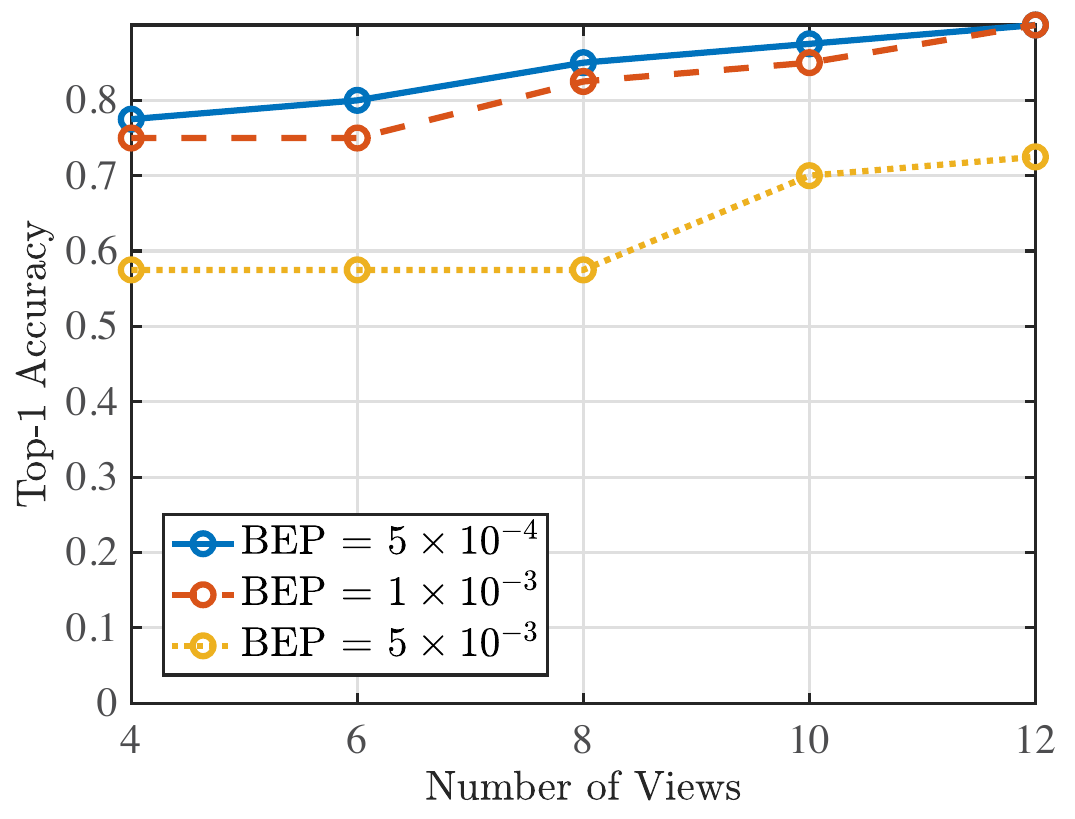}\label{Fig: DNN Accuracy vs Number of Views}}
\caption{Classification accuracy for DNN classifier (i.e., VGG-11) with real data (i.e., multi-view ModelNet40) for (a) single-view observation or (b) multi-view observation.}
\label{Fig: DNN experiment}
\end{figure}

\subsection{Robotic SemCom with DNN Classification}
The insights from analysis and design based on linear classification are validated using DNN models and real datasets.
To this end, the curves of top-1 accuracy are plotted in Fig. \ref{Fig: DNN experiment} for both the cases of single-view and multi-view classification against BEP and view number, respectively. 
Similar to its linear classification counterpart on Fig. \ref{Exp: Ultra-Low-Latency}, the classification margin is observed from Fig. \ref{Fig: DNN Accuracy vs BER} to exist in the current case and span the BEP range from 0 to about $10^{-3}$.
In other words, in the real world, the DNN model can tolerate significant channel distortion corresponding to BEP up to $10^{-3}$ without significantly compromising the accuracy.
On the other hand, the multi-view gain in classification margin is corroborated in Fig. \ref{Fig: DNN Accuracy vs Number of Views}.
Similar to the linear classification case in Fig. \ref{Fig: Accuracy vs Number of views}, the accuracy improvement rates are comparable for different BEPs.
The large drop in accuracy from BEP $=\{5\times10^{-4},1\times10^{-3}\}$ to BEP $=5\times 10^{-3}$ is due to the reason that the former two values are manageable by the classifier while the last value exceeds the limit of classifier robustness.

\begin{figure}[t!]
\centering
\includegraphics[width=0.7\columnwidth]{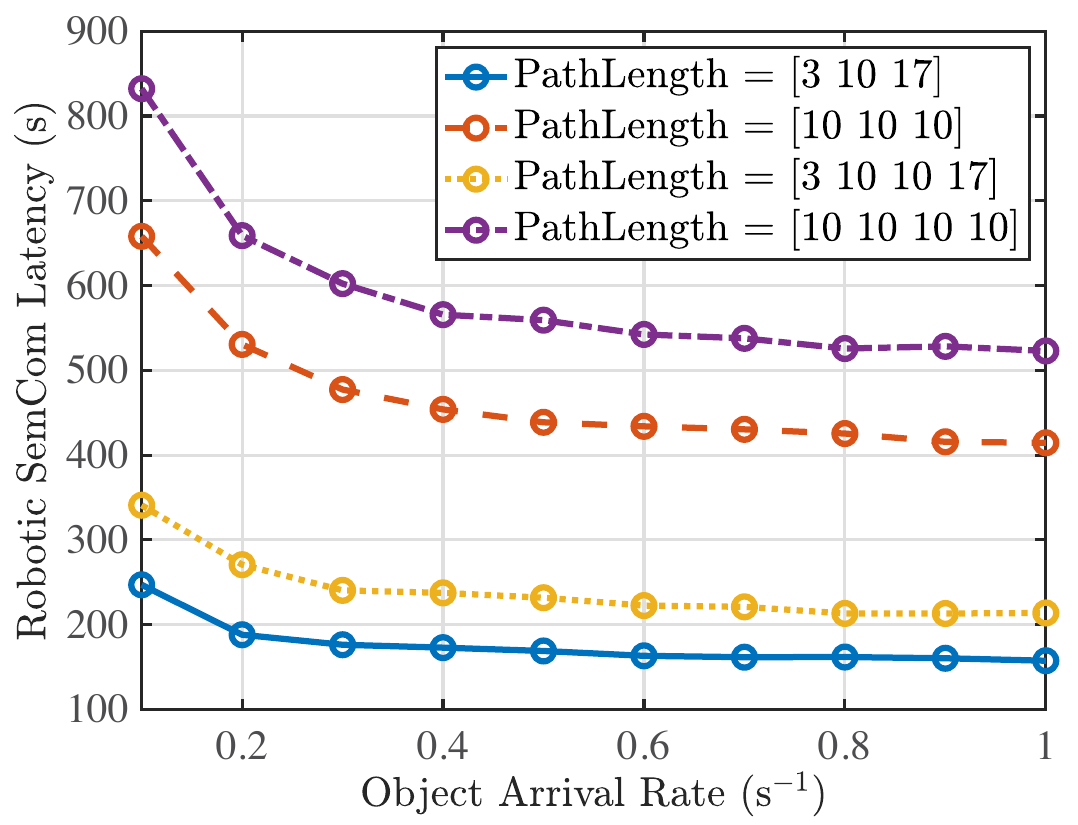}\label{Fig: Poission Latency}
\caption{Robotic SemCom latency as a function of the objects' arrival rate under different settings of the number of task-relevant KPs and corresponding path lengths.}
\end{figure}

Next, the curves of robotic exploration latency versus the object arrival rate are plotted in Fig. \ref{Fig: Poission Latency} for different settings of task-relevant KPs' length.
The robotic SemCom latency refers to the duration of the whole exploration process.
Firstly, a higher arrival rate is observed to reduce the SemCom latency due to faster robot mobility increases the likelihood of encountering task-relevant objects in a given interval. 
% It should be noted that when there is a low arrival rate, latency is mainly impacted by the time required to encounter objects. 
On the other hand, when the arrival rate is high, latency is predominantly affected by transmission. This explains why, in the regime of high arrival rate, the latency saturates even as the arrival rate continues to rise.
Secondly, by comparing the curves of different path-lengths, it is found that the exploration process can be completed more rapidly if there exists a short KP (e.g., with a length of 3 objects). Furthermore, given arrival rate, a greater number of task-relevant KPs indicates the presence of more objects in the environment. That increases the time required for path-hitting, namely locating one task-relevant KP.

\begin{comment}
\section{Concluding Remarks}
This work presents a knowledge-based SemCom framework for connected robotic intelligence systems. 
It introduces a robotic SemCom protocol and an ultra-low latency feature transmission scheme. 
The protocol identifies feasible KPs from a KG stored at edge server, enabling efficient object recognition through feature transmission. 
Our study addresses the issue of transmission errors by the robustness of classification margin and analyzes the impact of BEP on inference accuracy.  
In addition, the work proposes a round-robin mode system that leverages multi-view gain to enhance performance and presents an analysis of latency in the overall system. 
The proposed exploitation of knowledge based SemCom for robotic edge intelligence opens a new direction for designing 6G task-oriented communication techniques to support the connected machine intelligence at the edge. 
In the current context, there is potential for the advancement of the robotic SemCom framework by integrating advanced physical-layer techniques, e.g., massive MIMO and beamforming. 
In addition, extending the protocol to accommodate multiple robots through a traditional multi-access scheme and developing a suitable resource allocation algorithm is a noteworthy direction for exploration. Within this context, designing a cooperation scheduling protocol tailored to the needs of multiple robots presents an intriguing avenue. 
% where exploring the design of a cooperation scheduling protocol for multiple robots would be an intriguing aspect to consider.
\end{comment}

\section{Concluding Remarks}
In this paper, we presented a novel knowledge-based robotic SemCom framework, aiming to support robotic edge-AI systems where an edge server serves as a ``remote brain'' for the robot. The objective is to find a feasible KP on a large-scale KG to guide the robot in accomplishing a given task. To achieve this goal, we introduced a robotic SemCom protocol and an air-interface to support ULL-FT for data-intensive transmission. Furthermore, we explored two approaches to enhance the classification robustness, retransmission and multi-view, and validated their benefits when the classification margin is insufficient for counteracting channel distortion. Experimental results demonstrated the effectiveness of the proposed ULL-FT scheme in reducing communication latency while maintaining accurate feasible KP identification.

This study contributes to the ongoing research on SemCom in 6G networks and sets the stage for the widespread deployment of AI algorithms at the network edge, enabling advanced robotic edge intelligence systems. There are potential research topics for future research. These may include investigating more sophisticated KG models, developing advanced semantic matching techniques, incorporating advanced error control coding techniques, optimizing for multiple-access robotic systems and integrating the framework with other emerging technologies in 6G networks. Addressing these challenges will further enhance the performance and capabilities of the robotic  SemCom framework, leading to more efficient and intelligent robotic systems in future mobile networks.

% \vspace{-2mm}
\appendix

\subsection{Proof of Proposition \ref{proposition: margin}}\label{proof: margin}
Firstly, we introduce a lemma to facilitate our derivation: ``If $\mathbf{A}$ and $\mathbf{B}$ are two symmetric positive semi-definite matrices, then $\text{tr}(\mathbf{A}\mathbf{B})\geq0$ and $\text{tr}(\mathbf{A}\mathbf{B})\leq\text{tr}(\mathbf{A})\text{tr}(\mathbf{B})$."
The proof for this lemma is omitted for brevity. 
From the definition of Mahalanobis distance, we have
\begin{align}
    \sigma^2&=(\tilde{\mathbf{x}}-\bmu)^T(\mathbf{C}+\sigma^2_{\Delta}\mathbf{I})^{-1}(\tilde{\mathbf{x}}-\bmu)\nonumber\\
    &=\text{tr}\left\{(\mathbf{C}+\sigma^2_{\Delta}\mathbf{I})^{-1}(\tilde{\mathbf{x}}-\bmu)(\tilde{\mathbf{x}}-\bmu)^T\right\},\label{eqn: sigma}\\
    {\sigma'}^2&=(\tilde{\mathbf{x}}-\bmu)^T\mathbf{C}^{-1}(\tilde{\mathbf{x}}-\bmu)\nonumber\\
    &=\text{tr}\left\{\mathbf{C}^{-1}(\tilde{\mathbf{x}}-\bmu)(\tilde{\mathbf{x}}-\bmu)^T\right\}.
\end{align}
The difference of them gives
\begin{align}
    {\sigma'}^2\!-\!\sigma^2\!
    = \text{tr}\!\left\{\!\text{diag}\!\left\{\!\frac{\sigma_{\Delta}^2}{C_d(C_d+\sigma_{\Delta}^2)}\!\right\}\!(\tilde{\mathbf{x}}\!-\!\bmu)(\tilde{\mathbf{x}}\!-\!\bmu)^T\!\right\}\!.
\end{align}
Here, both $\text{diag}\left\{\frac{\sigma_{\Delta}^2}{C_d(C_d+\sigma_{\Delta}^2)}\right\}$ and $(\tilde{\mathbf{x}}-\bmu)(\tilde{\mathbf{x}}-\bmu)^T$ are symmetric positive semi-definite matrices. It is obvious that the diagonal matrix, $\max_{d}\left\{\frac{\sigma_{\Delta}^2}{C_d(C_d+\sigma_{\Delta}^2)}\right\}\mathbf{I}-\text{diag}\left\{\frac{\sigma_{\Delta}^2}{C_d(C_d+\sigma_{\Delta}^2)}\right\}$,
is still positive semi-definite, so that
\begin{align}\label{eqn: left}
    0&\leq\text{tr}\Big\{\left(\max_{d}\left\{\frac{\sigma_{\Delta}^2}{C_d(C_d+\sigma_{\Delta}^2)}\right\}\mathbf{I}-\text{diag}\left\{\frac{\sigma_{\Delta}^2}{C_d(C_d+\sigma_{\Delta}^2)}\right\}\right)\nonumber\\
    &\qquad\quad\times(\tilde{\mathbf{x}}-\bmu)(\tilde{\mathbf{x}}-\bmu)^T\Big\}\nonumber\\
    &=\max_d\left\{\frac{\sigma_{\Delta}^2}{C_d(C_d+\sigma_{\Delta}^2)}\right\}\text{tr}\left\{(\tilde{\mathbf{x}}-\bmu)(\tilde{\mathbf{x}}-\bmu)^T\right\}\nonumber\\
    &\qquad\quad-\text{tr}\left\{\text{diag}\left\{\frac{\sigma_{\Delta}^2}{C_d(C_d+\sigma_{\Delta}^2)}\right\}(\tilde{\mathbf{x}}-\bmu)(\tilde{\mathbf{x}}-\bmu)^T\right\}\nonumber\\
    &=\max_d\!\left\{\!\frac{\sigma_{\Delta}^2}{C_d(C_d\!+\!\sigma_{\Delta}^2)}\!\right\}\text{tr}\!\left\{\!(\tilde{\mathbf{x}}\!-\!\bmu)(\tilde{\mathbf{x}}\!-\!\bmu)^T\right\}-({\sigma'}^2\!-\!\sigma^2).
\end{align}
Then, we can derive the upper bound for the classification margin reduction as follows:
\begin{align}\label{eqn: right}
    \Delta\Phi=\Phi-\Phi'=\sigma'  -\sigma=\frac{{\sigma'}^2  -\sigma^2}{\sigma'+\sigma}\leq\frac{{\sigma'}^2  -\sigma^2}{2\sigma}.
\end{align}
Combining results \eqref{eqn: sigma}, \eqref{eqn: left} and \eqref{eqn: right} gives the upper bound:
\begin{align}
    \Delta\Phi&\leq\max_d \!\left\{\!\frac{\sigma_{\Delta}^2\sigma}{2C_d(C_d\!+\!\sigma_{\Delta}^2)}\!\right\}\frac{\text{tr}\!\left\{(\tilde{\mathbf{x}}\!-\!\bmu)(\tilde{\mathbf{x}}\!-\!\bmu)^T\right\}}{\text{tr}\!\left\{\!(\mathbf{C}\!+\!\sigma^2_{\Delta}\mathbf{I})^{-1}(\tilde{\mathbf{x}}\!-\!\bmu)(\tilde{\mathbf{x}}\!-\!\bmu)^T\right\}}\nonumber\\
    &=\max_d \left\{\frac{\sigma_{\Delta}^2\sigma}{2C_d(C_d+\sigma_{\Delta}^2)}\right\}\nonumber\\
    &\qquad\quad\times\frac{\text{tr}\left\{(\mathbf{C}+\sigma^2_{\Delta}\mathbf{I})(\mathbf{C}+\sigma^2_{\Delta}\mathbf{I})^{-1}(\tilde{\mathbf{x}}-\bmu)(\tilde{\mathbf{x}}-\bmu)^T\right\}}{\text{tr}\left\{(\mathbf{C}+\sigma^2_{\Delta}\mathbf{I})^{-1}(\tilde{\mathbf{x}}-\bmu)(\tilde{\mathbf{x}}-\bmu)^T\right\}}\nonumber\\
    &\leq\max_d \left\{\frac{\sigma_{\Delta}^2\sigma}{2C_d(C_d+\sigma_{\Delta}^2)}\right\}\text{tr}\left\{\mathbf{C}+\sigma^2_{\Delta}\mathbf{I}\right\}.
\end{align}
The derivation of the lower bound follows the similar procedures. This completes the proof.

% \vspace{-4mm}
\subsection{Proof of Lemma \ref{lemma: accuracy}}\label{proof: accuracy}
As shown in \eqref{eqn: score distribution}, the conditional distribution for the transmitted data score $s(\mathbf{x})$ is a Gaussian and thus the probability can be derived as follows:
\begin{align}
    &\Pr\left(s(\mathbf{x})s(\hat{\mathbf{x}})>0|\hat{\mathbf{x}}\right)\nonumber\\
    &=\frac{1}{\sqrt{2\pi\sigma_{\Delta}^2}}\int_{-\infty}^{|s(\hat{\mathbf{x}})|}\exp\left(-\frac{t^2}{2\sigma_{\Delta}^2}\right)dt\nonumber\\
    &=1-\frac{1}{\sqrt{2\pi}}\int_{\frac{|s(\hat{\mathbf{x}})|}{\sigma_{\Delta}}}^{\infty}\!\exp\!\left(\!-\frac{1}{2}u^2\!\right)du\nonumber\\
    &=1-Q\left(\frac{|s(\hat{\mathbf{x}})|}{\sigma_{\Delta}}\right),
\end{align}
where $Q(\cdot)$ is the Q-function.
This completes the proof.

% \vspace{-4mm}
\subsection{Proof of Proposition \ref{proposition: performance}}\label{proof: performance}
According to the law of total probability, we can derive the probability as follows:
\begin{align}
    &\Pr\left(s(\mathbf{x})s(\hat{\mathbf{x}})>0\mid\ell=0\right)\nonumber\\
    &=\int_{-\infty}^{\infty}\Pr\left(s(\mathbf{x})s(\hat{\mathbf{x}})>0\mid \hat{\mathbf{x}}\right)dF_{\hat{\mathbf{x}}}(\hat{\mathbf{x}})\nonumber\\
    &=1-\int_{-\infty}^{\infty}Q\left(\frac{|s(\hat{\mathbf{x}})|}{\sigma_{\Delta}}\right)f_{s(\hat{\mathbf{x}})}(s(\hat{\mathbf{x}}))ds(\hat{\mathbf{x}})\nonumber\\
    &=1-\int_{-\infty}^{\infty}Q\left(\frac{|t|}{\sigma_{\Delta}}\right)\frac{1}{\sqrt{2\pi\mathbf{w}^T(\mathbf{C}+\sigma_{\Delta}^2\mathbf{I})\mathbf{w}}}\nonumber\\
    &\qquad\qquad\qquad\times\exp\left(-\frac{(t-s(\bmu_0))^2}{2\mathbf{w}^T(\mathbf{C}+\sigma_{\Delta}^2\mathbf{I})\mathbf{w}}\right)dt,\label{eqn: integral}
\end{align}
where $F_{\hat{\mathbf{x}}}(\cdot)$ denotes the distribution function of $\hat{\mathbf{x}}$. Leveraging the Chernoff bound for Q-function, we have the inequality:
% $
%    Q\left(\frac{|s(\hat{\mathbf{x}})|}{\sigma_{\Delta}}\right)\leq\frac{1}{2}\exp\left(-\frac{s(\hat{\mathbf{x}})^2}{2\sigma_{\Delta}^2}\right).
% $
% Then, the integral in \eqref{eqn: integral} can be upper bounded by
\begin{align}
    &\int_{-\infty}^{\infty}Q\left(\frac{|t|}{\sigma_{\Delta}}\right)\exp\left(-\frac{(t-s(\bmu_0))^2}{2\mathbf{w}^T(\mathbf{C}+\sigma_{\Delta}^2\mathbf{I})\mathbf{w}}\right)dt\nonumber\\
    &\leq\int_{-\infty}^{\infty}\exp\left(-\frac{t^2}{2\sigma_{\Delta}^2}-\frac{(t-s(\bmu_0))^2}{2\mathbf{w}^T(\mathbf{C}+\sigma_{\Delta}^2\mathbf{I})\mathbf{w}}\right)dt\nonumber\\
    &=\sqrt{2\pi}\left(\frac{1}{\sigma_{\Delta}^2}+\frac{1}{\mathbf{w}^T(\mathbf{C}+\sigma_{\Delta}^2\mathbf{I})\mathbf{w}}\right)^{-\frac{1}{2}}\nonumber\\
    &\qquad\qquad\quad\times\exp\left(-\frac{s(\bmu_0)^2}{2(\sigma_{\Delta}^2+\mathbf{w}^T(\mathbf{C}+\sigma_{\Delta}^2\mathbf{I})\mathbf{w})}\right).
\end{align}
Substituting the result into \eqref{eqn: integral}, we have
\begin{align}
    &\Pr\left(s(\mathbf{x})s(\hat{\mathbf{x}})>0\mid\ell=0\right)\nonumber\\
    &\geq\!1\!-\!\sqrt{\!\frac{\sigma_{\Delta}^2}{\sigma_{\Delta}^2\!+\!\mathbf{w}^T\!(\mathbf{C}\!+\!\sigma_{\Delta}^2\mathbf{I})\mathbf{w}}}\exp\!\!\left(\!-\frac{s(\bmu_0)^2}{2(\sigma_{\Delta}^2\!+\!\mathbf{w}^T\!(\mathbf{C}\!+\!\sigma_{\Delta}^2\mathbf{I})\mathbf{w})}\!\right)\nonumber\\
    &=\!1\!-\!\sqrt{\!\frac{\sigma_{\Delta}^2}{(1\!+\!\mathbf{w}^T\!\mathbf{w})\sigma_{\Delta}^2\!\!+\!\mathbf{w}^T\!\mathbf{C}\mathbf{w}}}\exp\!\!\left(\!-\frac{s(\bmu_0)^2}{2(\!\sigma_{\Delta}^2\!(\!1\!+\!\mathbf{w}^T\!\mathbf{w}\!)\!+\!\mathbf{w}^T\!\mathbf{C}\mathbf{w}\!)}\!\!\right)\nonumber\\
    &=\!1\!-\!\sqrt{\!\frac{\sigma_{\Delta}^2}{2\sigma_{\Delta}^2\!+\!\mathbf{w}^T\!\mathbf{C}\mathbf{w}}}\exp\!\left(\!-\frac{s(\bmu_0)^2}{2(2\sigma_{\Delta}^2\!+\!\mathbf{w}^T\mathbf{C}\mathbf{w})}\right)\nonumber\\
    &=\!1\!-\!\sqrt{\!\frac{\sigma_{\Delta}^2}{2\sigma_{\Delta}^2\!+\!\mathbf{w}^T\!\mathbf{C}\mathbf{w}}}\exp\!\left(\!-\frac{\mathbf{w}^T\!\mathbf{C}\mathbf{w}}{2(2\sigma_{\Delta}^2\!+\!\mathbf{w}^T\!\mathbf{C}\mathbf{w})}\!\times\!\frac{s(\bmu_0)^2}{\mathbf{w}^T\!\mathbf{C}\mathbf{w}}\right)\nonumber\\
    &=\!1\!-\!\sqrt{\!\frac{\sigma_{\Delta}^2}{2\sigma_{\Delta}^2\!+\!\mathbf{w}^T\!\mathbf{C}\mathbf{w}}}\exp\!\left(\!-\frac{\mathbf{w}^T\!\mathbf{C}\mathbf{w}(\sigma\!+\!\Phi)^2}{2(2\sigma_{\Delta}^2\!+\!\mathbf{w}^T\!\mathbf{C}\mathbf{w})}\!\right)\!,
\end{align}
where we use the fact that $\mathbf{w}$ is a unit vector such that $\mathbf{w}^T\mathbf{w}=1$. 
Following the same procedures, the lower bound for the case of sampling feature vector $\mathbf{x}$ from the cluster centered at $\bmu_1$ has the same form except for replacing the score function $s(\bmu_0)$ by $s(\bmu_1)$.
Due to the uniform prior, we have the general expression of the lower bound and complete the proof.
% \begin{align}
%     &\Pr\!\left(s(\mathbf{x})s(\hat{\mathbf{x}})>0\right)\nonumber\\
%     &=\frac{1}{2}\Pr\!\left(s(\mathbf{x})s(\hat{\mathbf{x}})>0|\ell=0\right)+\frac{1}{2}\Pr\!\left(s(\mathbf{x})s(\hat{\mathbf{x}})>0|\ell=1\right)\nonumber\\
%     &=1-\sqrt{\!\frac{\sigma_{\Delta}^2}{2\sigma_{\Delta}^2+\mathbf{w}^T\mathbf{C}\mathbf{w}}}\exp\!\left(\!-\frac{(\sigma+\Phi)^2\mathbf{w}^T\mathbf{C}\mathbf{w}}{2(2\sigma_{\Delta}^2+\mathbf{w}^T\mathbf{C}\mathbf{w})}\right).
% \end{align}
% This completes the proof.
\subsection{Proof of Proposition \ref{proposition: multi-view}}\label{proof: multi-view}
If an object is observed for $m$ times, the aggregated feature vector is 
\begin{align}
    \overline{\hat{\bx}}=\frac{1}{m}\sum_{i=1}^m\hat{\bx}(i)=\frac{1}{m}\sum_{i=1}^m\bx(i)+\frac{1}{m}\sum_{i=1}^m\Delta\bx(i).
\end{align}
We introduce the notations $\overline{\bx}=\frac{1}{m}\sum_{i=1}^m\bx(i)$ and $\overline{\Delta\bx}=\frac{1}{m}\sum_{i=1}^m\Delta\bx(i)$. Then, we have
\begin{align}
    \overline{\bx}\sim\frac{1}{L}\sum_{\ell=1}^L\cN(\bmu_{\ell},\bC/m)\quad\text{and}\quad\overline{\Delta\bx}\sim\cN(\b0,\sigma_{\Delta}^2/m).
\end{align}
Given the aggregated feature vector $\overline{\hat{\bx}}$, the output score of the averaged feature vector $\overline{\bx}$ follows
\begin{align}
    s(\overline{\bx})\mid\overline{\hat{\bx}}\sim\cN(s(\overline{\hat{\bx}}),\sigma_{\Delta}^2/m).
\end{align}
Following the same procedures in the proof of Proposition 2 (see Appendix~\ref{proof: performance}), the correct classification accuracy can be derived as 
\begin{align}
    &\Pr\left(s(\overline{\hat{\bf x}})s(\overline{\mathbf{x}})>0|m\right)\nonumber\\
    &\geq
    \!1\!-\!\sqrt{\!\frac{\sigma_{\Delta}^2}{2\sigma_{\Delta}^2\!+\!\mathbf{w}^T\mathbf{C}\mathbf{w}}}\exp\!\left(-\frac{m\mathbf{w}^T\mathbf{C}\mathbf{w}(\sigma+\Phi)^2}{2(2\sigma_{\Delta}^2\!+\!\mathbf{w}^T\mathbf{C}\mathbf{w})}\right).
\end{align}
This completes the proof.

\subsection{Proof of Proposition \ref{proposition: retransmission}}\label{proof: retransmission}
Given error distribution $\Delta\bx\sim\l(\b0,\frac{\sigma_{\Delta}^2}{T}\bI\r)$, the classification accuracy of the received vector $\hat{\bx}$ is lower bounded by
\begin{align}
    &\Pr\left(s({\hat{\bf x}})s({\mathbf{x}})>0\right)\nonumber\\
    &\geq
    1\!-\!\sqrt{\frac{\sigma_{\Delta}^2}{2\sigma_{\Delta}^2\!+\!T\mathbf{w}^T\mathbf{C}\mathbf{w}}}\exp\!\left(\!-\frac{T\mathbf{w}^T\mathbf{C}\mathbf{w}(\sigma\!+\!\Phi)^2}{2(2\sigma_{\Delta}^2\!+\!T\mathbf{w}^T\mathbf{C}\mathbf{w})}\!\right)\!\!,
\end{align}
which comes from substituting $\sigma_{\Delta}^2/T$ for $\sigma_{\Delta}^2$ in Proposition 2.
To achieve the target classification accuracy $\xi$, the sufficient condition is to set the lower bound of $\Pr\left(s({\hat{\bf x}})s({\mathbf{x}})>0\right)$ equal to $\xi$. For ease of exposure, we denote
\begin{equation}\label{eqn: beta}
    \beta=\sqrt{\frac{\sigma_{\Delta}^2}{2\sigma_{\Delta}^2+T\mathbf{w}^T\mathbf{C}\mathbf{w}}},
\end{equation}
and thus the condition becomes
\begin{equation}
    \beta\exp\l((\beta^2-\frac{1}{2})(\sigma+\Phi)^2\r)=1-\xi.
\end{equation}
The solution to the equation gives
\begin{equation}\label{eqn: beta square}
    \beta^2=\frac{\cW\l(2(\sigma+\Phi)^2e^{(\sigma+\Phi)^2}(1-\xi)^2\r)}{2(\sigma+\Phi)^2},
\end{equation}
where $\cW(\cdot)$ is the Lambert $W$ function.
By substituting the expression in \eqref{eqn: beta square} into \eqref{eqn: beta}, we reach the result as follows:
\begin{equation}
    T=\frac{\sigma_{\Delta}^2}{\bw^T\bC\bw}\!\l(\!\frac{2(\sigma+\Phi)^2}{\cW\!\l(2(\sigma+\Phi)^2e^{(\sigma+\Phi)^2}(1-\xi)^2\r)}-1\!\r).
\end{equation}
The integer constraint for $T$ requires an additional ceiling function on it, and this completes the proof.

\bibliographystyle{IEEEtran}
\bibliography{Ref}
\end{document}